\documentclass[12pt,preprint]{aastex}

\slugcomment{Draft version Nov 20, 2010.}

\shorttitle{NLTE analyses of EHes}
\shortauthors{Pandey \& Lambert}

\begin{document}

\title{Neon and CNO Abundances for Extreme Helium Stars -- A Non-LTE Analysis}

\author{Gajendra Pandey}
\affil{Indian Institute of Astrophysics;
Bangalore,  560034 India}
\email{pandey@iiap.res.in}

\and

\author{David L.\ Lambert}
\affil{The W.J. McDonald Observatory, University of Texas at Austin; Austin,
TX 78712-1083}
\email{dll@astro.as.utexas.edu}

\begin{abstract}

A non-LTE (NLTE) abundance analysis was carried out for three
extreme helium stars (EHes): BD+10$^\circ$ 2179, BD$-9^\circ$ 4395, and
LS\,IV\,+6$^{\circ}$\,002, from their optical spectra with NLTE model
atmospheres.
NLTE TLUSTY model atmospheres were computed with 
H, He, C, N, O, and Ne treated in NLTE. 
Model atmosphere parameters were chosen from consideration of fits to
observed He\,{\sc i} line profiles and ionization equilibria of C and N ions.
The program SYNSPEC was then used to determine the 
NLTE abundances for Ne as well as 
H, He, C, N, and O. LTE neon abundances from Ne\,{\sc i} lines in 
the EHes: LSE\,78, V1920\,Cyg, HD\,124448, and PV\,Tel, are derived 
from published models and an estimate of the NLTE correction applied 
to obtain the NLTE Ne abundance. 

We show that the derived abundances of these key elements, including Ne, are
well matched with semi-quantitative predictions for the EHe resulting from
a cold merger (i.e., no nucleosynthesis during the merger) of a
He white dwarf with a C-O white dwarf. 

\end{abstract}


\clearpage

\section{Introduction}

The principal class of hydrogen-deficient supergiant stars  comprises three
subclasses
which in order of increasing but overlapping temperature intervals from
coolest to hottest are  the H-deficient carbon stars (HdC), the R Coronae Borealis stars (RCB),
and the extreme helium stars (EHe).
A common supposition is that the three subclasses are related in terms of origin and
evolution. The origin of these very rare stars has long been disputed but it now
seems likely that the majority are formed through a merger of a He white
dwarf with a C-O white dwarf, the so-called double degenerate (DD) 
scenario \citep{webb84,ibandtutu84,saio02}. Others may be the result of a 
final He-shell flash in a post-AGB star, the so-called final
flash (FF) scenario \citep{iben83,iben96,herwig2001,bloc2001}.

Much of the evidence for deciding whether  HdC, RCB, or EHe stars
come from the DD or FF
scenario (or neither)  depends on comparison of the
observed chemical composition with predictions by the two scenarios. It is in this
context that we present in this paper a non-LTE (NLTE) analysis of the neon abundance of
a sample of EHe stars where Ne\,{\sc i} lines are prominent in optical spectra;
neon is detectable in EHe stars, the warmer RCBs but not the HdCs.
(The NLTE analyses are extended here to He, C, N, and O lines.) 
%Surface neon 
%enrichment is likely in
%the DD scenario; details of the enrichment process are uncertain as we note
%later in the paper. Enrichment in the FF scenario is deemed unlikely.
%Thus, the Ne abundance is a potential discriminator between the DD and FF hypotheses.

If reliable Ne abundances can be provided for EHes and RCBs, neon will join other
abundances as clues to the origins and evolution of the H-deficient
supergiants. In addition to the obvious importance of C, N, and O elemental
abundances, one may now note a variety of other abundance anomalies
peculiar to these supergiants including, for example, the presence of lithium in
a subset of RCBs and one HdC \citep{asp00,raolamb96}, the large overabundance of fluorine 
in EHes and RCBs \citep{panF2006,pan08}, high concentrations of 
$^{18}$O (relative to $^{16}$O) in HdCs and cool RCBs \citep{clay2005,clay2007,garc2009,garc2010}, extraordinary
high Si/Fe and S/Fe ratios in the `minority' RCBs \citep{raolamb96}.

If neon is to provide an effective addition to the list of abundance anomalies, its abundance must be
determined reliably and, in this regard, the primary consideration 
would appear to be  an adequate
treatment of NLTE effects in the formation of the observable neon lines. Realization
that NLTE effects are considerable for optical Ne\,{\sc i} lines arose from pioneering
calculations by \citet{aur73} for normal B-type stars with effective
temperatures of around 20,000 K. These authors showed that the Ne abundance derived by
accounting for NLTE effects was about a factor of five less than that given by LTE. Not
only was this the first result showing major NLTE effects on abundances for hot
stars but the NLTE Ne abundance was shown to be in good agreement with that for H\,{\sc ii} regions
as derived from emission lines. The origin of the marked NLTE effects is discussed
by Auer \& Mihalas. A key ingredient is that the ultraviolet Ne\,{\sc i} resonance
lines are optically thick --- see a concise discussion by \citet{chl06} who
report on modern calculations of Ne NLTE effects as applied to B stars in the
Orion Association. Given that the ultraviolet resonance lines may be
similarly optically thick in atmospheres of EHe stars, it became apparent that addition of
neon to the list of referees between DD and FF scenarios would require evaluation of the
NLTE effects on the observable neon lines.

In the following sections, we successively describe our optical spectra, the NLTE
calculations including a sanity check involving our analysis of  normal B stars
previously discussed by \citet{chl06} and \citet{mandb08},
our abundance analyses of seven EHes, a discussion of the DD scenario with
a comparison of semi-quantitative predictions with the observationally-based
abundances of He, C, O, and Ne as well as remarks on abundances not
determinable for EHes (e.g., Li, $^{18}$O, and F). This comparison is followed
by remarks on the FF scenario and a few concluding remarks.

\section{Observations}

High-resolution optical spectra of BD+10$^\circ$ 2179, BD$-9^\circ$ 4395 and V1920\,Cyg
were obtained on 24 January 1998, 16 June 2000, and 25 July 1996,
respectively, at the coud\'e focus of the W.J. McDonald Observatory's
Harlan J. Smith 2.7-m telescope with the Robert G. Tull cross-dispersed
echelle spectrograph \citep{tull95} at a resolving power of
R=60,000 except for BD$-9^\circ$ 4395's spectrum acquired at R=40,000.
These spectra with R=60,000 were previously described by \citet{pan2006}.
Additional spectra of BD$-9^\circ$ 4395 were obtained on 18, 22, and 25 July 2002 
at the W.J. McDonald Observatory's Otto Struve 2.1-m telescope with the Sandiford 
Caasegrain echelle spectrograph \citep{McCar93} at a resolving
power of R=40,000. The spectrum of LSE\,78  was acquired with the 
Cassegrain echelle spectrograph at CTIO, and the observations are 
described in \citet{panred2006}. 
Finally, spectra at R=30,000 of the
two southern EHes -- PV\,Tel and HD\,124448 --
are from the Vainu Bappu Observatory and the fiber-fed cross-dispersed 
echelle spectrometer \citep{rao04,rao05b} at the 2.34-m telescope.
The Image Reduction and Analysis Facility (IRAF) software package was
used to reduce all spectra.\footnote{The IRAF software is
distributed by the National Optical Astronomy Observatories under contract
with the National Science Foundation}

Sample wavelength intervals in Figures 1 and 2 include one or two of the Ne\,{\sc i}
lines with the EHes ordered from top to bottom in order
of decreasing effective temperature.
All spectra are aligned to the rest wavelengths of well-known lines.
Inspection of the figures shows that the line profiles are not always
symmetric. Asymmetries obviously present in the case of LSE\,78 and V1920\,Cyg
are most probably due to atmospheric pulsations. In the case of V1920\,Cyg,
another observation on 1996 July 26, one day following the spectrum illustrated
in Figures 1 and 2, showed symmetric line profiles with no appreciable change in the
equivalent widths of lines. The equivalent width change translates to an abundance
change of less than 0.1 dex. In the case of BD$-9^\circ$ 4395, emission components
may appear and disappear. Variable photospheric spectra for EHes are 
common \citep{jef08} with V1920\,Cyg representative of the variability 
and BD$-9^\circ$ 4395 as an extreme example. Nonetheless, we assume that models constructed
with classical assumptions (plane parallel layers in hydrostatic equilibrium)
are adequate for our purpose. 

\section{NLTE Atmospheres and Abundances}

Our calculations use codes developed by Hubeny and colleagues, that is the
program TLUSTY for calculating LTE and NLTE model atmospheres \citep{hub88,hl95}
and the spectrum synthesis code SYNSPEC \citep{hub95}. 
In exercising these codes, we adopt the atomic data and model atoms provided on the 
TLUSTY homepage\footnote{http://nova.astro.umd.edu/index.html} \citep{lh07,lh03}.

The suite of codes was imported to Bangalore and run by one of us (GP) on a local
computer. Before proceeding to construct and apply H-deficient model atmospheres, our imported
codes were tested for normal B-type stars. In particular, we computed a NLTE model
atmosphere for HD 35299, a normal B star in the Orion sample for which \citet{chl06} derived
NLTE Ne abundances. The NLTE TLUSTY model was computed for the stellar parameters adopted by
Cunha et al.: $T_{\rm eff}$ = 24000K, $\log g$= 4.25 cgs and a microturbulence of 2 km s$^{-1}$
and solar abundances.  The $gf$-values for the Ne\,{\sc i} lines
are taken from \citet{seat98} who showed that Opacity Project theoretical $gf$-values
are in very good agreement not only with theoretical calculations of comparable sophistication
but also with experimental determinations.
Then, the NLTE Ne abundances were computed using the TLUSTY model
by matching the observed equivalent width of Ne\,{\sc i} lines with
NLTE predictions from running the SYNSPEC code. Observed equivalent widths were kindly
provided by Dr. Katia Cunha (private communication) for the eight lines measured
by them. Our NLTE Ne abundance for the eight lines is $\log$(Ne)= 8.20$\pm$0.08
in good agreement with the value of 8.18 given by Cunha et al. 
This agreement over NLTE Ne abundances is taken as evidence that our 
implementation of the TLUSTY-related codes was successful.

As a second check, we analysed Ne\,{\sc i} and Ne\,{\sc ii} lines in $\beta$ CMa.
\citet{mandb08} analyzed 7 Ne\,{\sc i} and 4 Ne\,{\sc ii} lines in this star.
Morel \& Butler compute the NLTE Ne abundances for a LTE Kurucz model with the
parameters: $T_{\rm eff}$ = 24000K, $\log g$= 3.5 cgs and
a microturbulence of 14 km s$^{-1}$.
We computed
a NLTE TLUSTY model  for these stellar parameters. 
The $gf$-values and the measured equivalent widths of the Ne\,{\sc i} and the 
Ne\,{\sc ii} lines were taken from Morel \& Butler. NLTE Ne abundances were 
computed using the TLUSTY atmosphere and model atoms 
by matching the measured width of Ne\,{\sc i} and Ne\,{\sc ii} lines with 
NLTE predictions from the SYNSPEC code. Our NLTE Ne abundance for the seven Ne\,{\sc i}
lines is $\log$(Ne)= 7.89$\pm$0.09 in agreement with the value of 7.89$\pm$0.04
given by Morel \& Butler. For the three Ne\,{\sc ii} lines, our NLTE Ne abundance 
is $\log$(Ne)= 8.16$\pm$0.16 where the value of 7.89$\pm$0.06 
is given by Morel \& Butler. 
Note that, one of the four Ne\,{\sc ii} lines returns a higher abundance  and was
not included in calculating our mean abundance. The different Ne abundances from
Ne\,{\sc ii} lines is noted and may arise from the use of different
models (TLUSTY NLTE versus Kurucz LTE) and the use of different model Ne atoms.
These checks on published NLTE Ne abundances are taken as evidence that our
implementation of the TLUSTY-related codes was successful.

A small grid of NLTE TLUSTY model atmospheres for EHe stars
was computed for $T_{\rm eff}$ from 15,000 K to 31,000 K 
and surface gravities $\log g$= 2.35 to 4.3.
The abundances adopted for the grid were representative of the LTE
abundances given by \citet{pan2006}. In particular, the C/He ratio was
assume to be 1 per cent.
Sample models for H/He=0.1 and 0.0001 showed that the derived abundances 
of neon and other elements are insensitive to the H abundance in this range.

\section{NLTE Abundance Analyses}

\subsection{BD +10$^\circ$ 2179}

An extensive LTE abundance analysis of BD+10$^\circ$ 2179 was reported by \citet{pan2006}
from optical and ultraviolet spectra. Abundances were obtained for 18 elements from H
to Zn but neon was not included. Here, we present a NLTE model atmosphere
redetermination of the He, C, N, and O abundances and the first determination of the
Ne abundance. The star's atmospheric parameters are reassessed using NLTE
atmospheres and NLTE line formation for He, C, N, O and Ne lines.

Optical lines of He\,{\sc i}, C\,{\sc i}-{\sc iii}, N\,{\sc ii}, O\,{\sc ii}
and Ne\,{\sc i} are used. Details about  these lines except for Ne\,{\sc i}
are taken from Table 2 of \citep{pan2006}. Details include a line's $gf$-value and
the reference to the source of that value, its lower excitation potential ($\chi$), 
and information on the line's Stark and radiative damping constants. 
Values from 2006 are adopted here in full. For Ne which is not in the 2006 Table, 
we adopt the $gf$-values from \citet{seat98}, as noted above.
Table 1 of this paper lists the chosen lines of C, N, O, and Ne where the equivalent
widths of Ne\,{\sc i} lines were measured off the spectrum used for the 2006 analysis.

Atmospheric parameters are obtained by the  procedures used for the 2006 LTE
analysis  but using NLTE TLUSTY model
atmospheres and NLTE line formation using the TLUSTY model atoms.
The microturblence is provided from the usual requirement that the
abundance from lines of a given species be independent of a line's equivalent
width: we use the C\,{\sc ii} lines. The effective temperature and surface gravity
are found from intersecting loci in the ($T_{\rm eff}$, $\log g$) plane with loci
provided by fits to He\,{\sc i} line profiles, and the ionization equilibria among
C$^\circ$, C$^+$, and C$^{2+}$. The LTE analysis is repeated but this time with
TLUSTY LTE model atmospheres instead of models from the code
STERNE \citep{jeff2001}.

Figure 3 illustrates the determination of the microturbulence from C\,{\sc ii} lines. A value of
7.5 km s$^{-1}$ is adopted. Although this value is for a particular model ($T_{\rm eff}$=17000K,
$\log g$= 2.5), the result is insensitive to the model choice. 
%The LTE value was also
%7.5 km s$^{-1}$ but a value of 6.5 km s$^{-1}$ was adopted by \citet{pan2006} because the
%N\,{\sc ii} lines indicated a value of 6 km s$^{-1}$. The effect on the abundance analysis
%of a choice between 6.5 or 7.5 km s$^{-1}$ is very small because our line list contains weak lines.

Sample theoretical NLTE line profiles and the observed profile of the  He\,{\sc i} 4471\AA\ line
are shown in Figure 4 for a model with 
an effective temperature of 16,375K and a surface gravity of 2.45 g cm$^{-2}$ and with
microturbulence and 
rotational broadening included (see \citet{pan2006}). The best-fitting theoretical
profile ($\log g$= 2.45) provides one point on the $T_{\rm eff} - \log g$ locus. 
The He\,{\sc i} lines at 
4009, 4026, and 4387 \AA\ lines were similarly analysed. The helium model atoms and
line broadening coefficients are from TLUSTY. 
Using the TLUSTY grid of model atmospheres, the loci were mapped out. The four loci
are shown in Figure 5 and are almost coincident.

Loci representing ionization equilibrium are provided from the requirements that
(C\,{\sc i}, C\,{\sc ii}), (C\,{\sc ii},C\,{\sc iii}), and (C\,{\sc i},C\,{\sc iii})
provide the same C abundance.

Figure 5 shows the several loci. Their intersection suggests that the best NLTE model
has $T_{\rm eff}$=16375$\pm$250K and $\log g$ = 2.45$\pm$0.2.  The best LTE TLUSTY model
with the LTE line analysis gives a best model with $T_{\rm eff}=17000$K and
$\log g$= 2.60. This LTE model differs a little from that 
adopted in the 2006 LTE analysis of the optical lines where
loci representing  ionization equilibria for (Si\,{\sc ii},Si\,{\sc iii}), (S\,{\sc ii},S\,{\sc iii}),
and (Fe\,{\sc ii},Fe\,{\sc iii}) were also considered. The 2006 LTE analysis gave a
model with $T_{\rm eff}$= 16400$\pm$500K and $\log g$= 2.35$\pm$0.2 cgs.

Abundances for C, N, O, and Ne are given in Table 1. Mean abundances and their standard deviations
are listed for both the NLTE and LTE TLUSTY analyses. 
Abundances are given as $\log\epsilon(X)$ 
and normalized to $\log \Sigma$$\mu_X \epsilon(X)$ = 12.15 where $\mu_X$ is the atomic weight of element X.
%$\log\Sigma\mu_X\epslion$(X)=12.15 where $\mu_X$ is the atomic weight of element X.
The NLTE abundance errors arising from uncertainties in the atmospheric parameters
are estimated by considering changes of $\Delta T$= $\pm$250K, $\Delta\log g$=$\pm$0.2 cgs, and
$\Delta\xi$ = $\pm$1 km s$^{-1}$. The rms errors in the abundances from C\,{\sc i}, C\,{\sc ii},
C\,{\sc iii}, N\,{\sc ii}, O\,{\sc ii}, and Ne\,{\sc i} are 0.22, 0.03, 0.18, 0.08, 0.12, and 0.10, 
respectively, with a negligible contribution from the microturbulence. The C/He ratio is 0.6\% 
but a ratio of 1\% was assumed in construction of the NLTE model. Recomputation
of the model for C/He=0.6\% results in negligible changes to the abundances in Table 1.
Abundance uncertainties are similar for the LTE analysis.

With the exception of  H\,{\sc i}, C\,{\sc iii} and the Ne\,{\sc i} lines, the
introduction of NLTE for the model atmosphere and line analysis has a
minor effect on the derived abundances. The mean abundance differences in dex in the sense
(LTE -- NLTE) are 0.07 (C\,{\sc i}), 0.04 (C\,{\sc ii}), $-$0.07 (N\,{\sc ii}),
$-$0.26 (O\,{\sc ii}), and 0.81 (Ne\,{\sc i}). 

The H\,{\sc i} lines show similar NLTE effects (LTE -- NLTE) across the lines. 
The difference in abundance (LTE -- NLTE) is about 0.33 dex. Note that, 
the NLTE/LTE abundance from H$\beta$ down the sequence decreases by about 0.3 dex.

The C\,{\sc iii} lines represent a fascinating issue in line formation. In the
LTE analysis, the 4186.9\AA\ 40 eV line gives an abundance that is
0.6 dex greater than that from the 4650\AA\ triplet which provides a more
plausible abundance. In NLTE, however, the
abundance discrepancy is reversed: the 4186\AA\ line gives a plausible abundance that is
0.7 dex less than that from the triplet. 
\citet{nieva2008} state that the sense of this reversal is expected according to their
calculations for normal B stars. The magnitude of the NLTE 
effects and the failure of our calculations to provide consistent NLTE
abundances suggests that the C\,{\sc iii} be given lower weight in the analysis.

There are small and unimportant
 differences between the 2006 LTE abundances and those in Table 1. Such
differences arise from a combination of factors: the model atmosphere codes are
different, and the derived atmospheric parameters are different. The differences in dex in
the sense (TLUSTY -- STERNE) are 0.12 (C\,{\sc i}), 0.04 (C\,{\sc ii}), 0.14 (N\,{\sc ii}),
0.18 (O\,{\sc ii}) and 0.04 (Ne\,{\sc i}).
 
%Note that Table 1 shows a large   for analysing  H\,{\sc i} lines, SYNSPEC code used an approximate Stark 
%broadening treatment \citep{hhl1994,hl95}.

\subsection{BD$-9^\circ$ 4395}

This star's spectrum contains  absorption lines with variable profiles and variable
emission lines
mainly from He\,{\sc i}, C\,{\sc ii}, and Si\,{\sc ii} transitions. 
These emission lines have been attributed to a shell or extended atmosphere.
An extensive library of optical and  ultraviolet spectra of BD$-9^\circ$ 4395 was discussed by
\citet{jefheb92} who undertook an
abundance analysis using absorption lines  
drawn from a mean optical spectrum. Their LTE analysis led to the atmospheric parameters:
$T_{\rm eff}$=22700$\pm$1200 K, $\log g$=2.55$\pm$0.10, and $\xi$ = 20$\pm$5 km s$^{-1}$. 
In addition to the line broadening from the high microturbulence and line profile variations,
the line profiles suggested the star may be rotating at about 40 km s$^{-1}$. 

Our high-resolution optical spectra confirm the characteristics described by Jeffery \& Heber. 
We measure equivalent widths off our spectra. Most of the measured equivalent widths are
from the 16 June 2000 spectrum. These measured equivalent widths are in fair agreement
with those measured off the spectra obtained on other dates.

Our abundance analysis follows the method discussed in the previous section.
Details about the majority of the  lines are taken from \citep{pan2006} with   
information on other lines of C\,{\sc ii-iii}, N\,{\sc ii-iii}, O\,{\sc ii},
and Ne\,{\sc ii} from the NIST database \\
(http://physics.nist.gov/PhysRefData/ASD/lines\_form.html).
The source of $gf$-values for Ne\,{\sc i} lines is as given in Section 4.1.

The O\,{\sc ii} lines confirm the high microturbulence with our
NLTE analyses giving $\xi$=17.5$\pm$5 km s$^{-1}$. This value is not sensibly
different from the 20 km s$^{-1}$ obtained by Jeffery \& Heber. The
microturbulence is somewhat higher than found for most other EHe stars 
and indicates supersonic atmospheric motions.

The He\,{\sc i} lines
are moderately sensitive to gravity. As clearly noted by Jeffery \& Heber, emission
affects the He\,{\sc i} profiles to differing degrees. For example, 
the 5876\AA\ line is in emission. Observed profiles of the 4143\AA\ and 4387\AA\
line are shown in Figure 6 with predicted NLTE profiles for a NLTE
atmosphere of $T_{\rm eff}$=24300K and three different surface gravities.
The predicted profiles have been convolved with a (Gaussian) profile with a FWHM of
40 km s$^{-1}$ to represent the  projected rotational velocity suggested
by Jeffery \& Heber. The chosen lines are those least affected by emission \citep{jefheb92}. 
There may be indications that weak emission contaminates the red wing and,
perhaps, the line core. LTE profiles shown by Jeffery \& Heber predict less deep cores
than  the observed profiles; the NLTE profiles reproduce the
line cores more closely than LTE profiles.

Ionization equilibria C\,{\sc ii}/C\,{\sc iii}, and N\,{\sc ii}/N\,{\sc iii} provide 
two loci in the ($T_{\rm eff}, \log g$) plane (Figure 7). 
Inspection of this figure suggests a solution with 
$T_{\rm eff}$ = 24300$\pm$700 K and $\log g$ = 2.65$\pm$0.20
cgs where we give equal weight to the C and N ionization equilibria.  
This effective temperature is 1600K hotter than
estimated by Jeffery \& Heber. The difference is partly accounted for by the fact that the
earlier (LTE) analysis included loci representing ionization equilibrium for Si\,{\sc ii}/Si\,{\sc IV}
and S\,{\sc ii}/S\,{\sc iii} and these loci of similar slope to the C and N loci fell about 1000K
to lower temperatures. Final abundances for our adopted model are given in Table 2.
Mean abundances and their standard deviations are given. The rms uncertainties arising from  the
estimated uncertainties of the atmospheric parameters are 0.05 (C\,{\sc ii}), 0.16 (C\,{\sc iii}),
0.08 (N\,{\sc ii}), 0.20 (N\,{\sc iii}), 0.02 (O\,{\sc ii}), 0.08 (Ne\,{\sc i}), and 0.16 (Ne\,{\sc ii}).

The LTE abundances in Table 2 were computed from a TLUSTY LTE model
atmosphere with  
model parameters  ($T_{\rm eff}$, $\log g$, $\xi$)=(24800, 2.85, 23.0).
Line by line LTE
abundances including the mean abundance
and the line-to-line scatter are given in Table 2.
These LTE abundances are quite similar to those reported by Jeffery \& Heber from
a different line list with different atomic data, 
a different model chosen from a different grid
of LTE atmospheres: the differences in dex in the sense (TLUSTY -- JH) are
0.22 (C\,{\sc ii}), $-0.35$ (C\,{\sc iii}), 0.03 (N\,{\sc ii}), $-0.01$ (N\,{\sc iii}),
0.05 (O\,{\sc ii}), 0.02 (Ne\,{\sc i}), and $-0.13$ (Ne\,{\sc ii}).

Corrections for NLTE effects in the sense (LTE -- NLTE) are $-0.34$ (H\,{\sc i}), 0.11 (C\,{\sc ii}),
$-0.07$ (C\,{\sc iii}), 0.32 (N\,{\sc ii}), 0.37 N\,{\sc iii}, $-0.09$ (O\,{\sc ii}),
0.60 (Ne\,{\sc i}), and $-0.01$ (Ne\,{\sc ii}) in dex. In the case of C and Ne, the two
stages of ionization treated in NLTE give consistent abundances  but do not in
LTE. Also, noteworthy is that the C\,{\sc iii} lines treated in NLTE give fairly
consistent results but this was not the case for BD$+10^\circ$ 2179. 

The NLTE correction (LTE -- NLTE) for H\,{\sc i}
is about $-0.34$ dex, a reversal in the NLTE correction that
was provided by the analysis of BD$+10^\circ$ 2179. It appears that
the NLTE correction (LTE -- NLTE) is mainly a function of effective temperature 
as these stars are of similar surface gravity.

\subsection{LSE\,78, V1920\,Cyg, HD\,124448, and PV\,Tel}

Neon abundances for this quartet are estimated by applying corrections to the
LTE Ne abundances from Ne\,{\sc i} lines
based on the NLTE calculations computed for model
atmosphere grids computed for BD$+10^\circ$ 2179 and BD$-9^\circ$ 4395. 
NLTE Ne abundances  for LSE\,78 and V1920\,Cyg are estimated by interpolation
in the grids of computed NLTE corrections
but for HD\,124448 and PV\,Tel an extrapolation is required.  
%The reference LTE Ne abundances are computed using the LTE STERNE model
%atmospheres used for our 2006 abundance analyses which did not include
%Ne. 
Neon LTE abundances are computed with the LTE models and
the Armagh LTE code SPECTRUM
\citep{jefheb92,jeff2001}.
In Tables 3, 4, 5, and 6, we give line by line LTE neon
abundances including the mean abundance, and the line-to-line scatter. 
%{\bf Note that, the line profiles of V1920\,Cyg change from
%one day to the next, being asymmetric and symmetric, respectively.
%However, the abundance determination from the measured equivalent widths of these
%profiles is not affected by not more than about 0.1 dex.}
%Inferred NLTE Ne abundances estimated from the mean LTE abundances
%are given in Table 8. 
The estimated NLTE corrections
to the LTE neon abundances of LSE\,78, V1920\,Cyg, HD\,124448, and PV\,Tel are
0.73, 0.8, 0.8, and 0.88, respectively. 

For LSE\,78 and V1920\,Cyg, the LTE Ne abundance is independent of a Ne\,{\sc i}'s line
equivalent width when the microturbulence from the 2006 paper is
adopted.  For HD 124448, two weak Ne\,{\sc i} lines provide the abundance and the
adopted value of the microturbulence is unimportant. 
In the case of PV\,Tel, the only
Ne\,{\sc i} lines available from our spectra are strong and the microturbulence from
the 2006 paper gives a Ne abundance that is a function of a line's equivalent
width,  a trend that may be removed by increasing the adopted value of the
microturbulence from the 15$\pm4$ km s$^{-1}$ found in 2006 from
optical N\,{\sc ii} and S\,{\sc ii} lines to 25 km s$^{-1}$ and then the
LTE Ne abundance is 8.53$\pm0.08$ (Table 6). An estimated NLTE
correction of 0.9 dex gives the NLTE Ne abundance of 7.6. 

\subsection{LS\,IV\,+6$^{\circ}$\,002}

Abundance analysis of LS\,IV\,+6$^{\circ}$\,002 was done by
\citet{jeff98} using absorption line equivalent 
widths drawn from the optical spectrum. This LTE analysis led to the atmospheric parameters:
$T_{\rm eff}$=31800$\pm$800 K, $\log g$=4.05$\pm$0.10, and $\xi$ = 9$\pm$1 km s$^{-1}$. 
This is the hottest star in our sample with Ne\,{\sc ii} but not Ne\,{\sc i} lines
in its spectrum.

Here \citet{jeff98}'s equivalent width have
been reanalyzed using our $gf$-values from \citep{pan2006} and the NIST database.
Two sets of model atmospheres are
considered: NLTE/TLUSTY and LTE/TLUSTY.
Analyses of the C\,{\sc iii}, N\,{\sc ii}, and O\,{\sc ii} lines confirm the microturbulence 
obtained by Jeffery with our
NLTE and LTE analyses giving $\xi$ about 9 km s$^{-1}$.
Ionization equilibria C\,{\sc ii}/C\,{\sc iii}, and N\,{\sc ii}/N\,{\sc iii} provide 
two loci in the ($T_{\rm eff}, \log g$) plane (Figure 8). The He\,{\sc i} 4471\AA\ line
that is moderately insensitive to gravity provides another locus. 

Inspection of Figure 8, produced by adopting NLTE/TLUSTY models, suggests a solution
with $T_{\rm eff}$=30000$\pm$800 K and $\log g$=4.10$\pm$0.15
cgs. The He\,{\sc ii} 4686\AA\ line suggests an
effective temperature about 1000 - 2000 K hotter. Here we give more weight to the
C and N ionization equilibria, and the locus provided by the He\,{\sc i} 4471\AA\ line.
This effective temperature is 2000K cooler than
estimated by Jeffery. Final abundances for our adopted model are given in Table 7.
Mean abundances and their standard deviations are given.
Corrections for NLTE effects in the sense (LTE -- NLTE) in dex are as
follows: $-0.33$ (H\,{\sc i}), $0.13$ (C\,{\sc ii}),
$-0.36$ (C\,{\sc iii}), $-0.30$ (N\,{\sc ii}), 0.25 N\,{\sc iii}, 0.14 (O\,{\sc ii}),
and $-0.02$ (Ne\,{\sc ii}).

The LTE abundances in Table 7 were computed from a TLUSTY LTE model
atmosphere.
The best TLUSTY LTE model parameters are ($T_{\rm eff}$, $\log g$, $\xi$)=(32000, 4.20, 9.0).
Note that, no weight is given to the C ionization equilibrum suspecting
departures from LTE.
Line by line LTE abundances including the mean abundance
and the line-to-line scatter are given in Table 7.
These LTE abundances are quite similar to those reported by Jeffery from
a different line list with different atomic data, 
a different model chosen from a different grid
of LTE atmospheres. 

Our NLTE Ne abundance in Table 7 is based on Kurucz $gf$-values for the Ne\,{\sc ii}
lines. The NLTE corrections
for these Ne\,{\sc ii} lines are small being typically 0.02 dex in the sense that the
NLTE abundance is  higher than the LTE value. 
Jeffery's 1998 LTE Ne abundance is based on $gf$-values that are systematically smaller
than our adopted values with a mean difference of 0.7 dex. Thus our LTE Ne abundance is
0.7 dex lower than Jeffery's value of 9.33.

\section{Interpreting the neon and CNO abundances}

\subsection{The context}

Knowledge of the chemical composition of EHe stars has become
more complete in recent years. 
Neon adds a new constraint on proposed
origins for these H-deficient stars. In order to exploit
this probe fully, the Ne abundance must be considered with the
reported abundances of other elements in EHe stars. 
A summary of He, C, N, O, Ne, and Fe abundances is given in Table 8. NLTE
abundances are given for C, N, O, and Ne but not for Fe.
For this review, we rely heavily
on our earlier analysis \citep{pan2006}. Two of our seven stars were not included
in the 2006 analysis and two from that analysis are not discussed here. For selected
points below, we comment on abundance aspects for other EHes with an abundance analysis. 

The following appear to be key points:

Hydrogen: Hydrogen is truly a trace element with depletion
factors of 10$^3$ or greater.

Carbon/Helium ratio: The C/He ratio runs from about 0.3\% to
1.0\% by number for the stars in Table 8. Five cool EHe stars \citep{pan2001,panred2006}
and the two other EHe stars discussed by \citep{pan2006} also fall within this range. 
%The hot RCB DY\,Cen with C\He=1\% \citep{jef1993} which might
%be deemed a EHe falls in the range too. 
Two known EHe stars fall well
below the range: V652\,Her with C/He=0.004\% \citep{jeff99} and
HD\,144941 with C/He=0.002\% \citep{har1997}. One supposes that the scenario
accounting for the stars in Table 8 and others with a similar C/He ratio will
need major revision to account for V652\,Her and HD\,144941.

Nitrogen: The N abundance is generally  equal to
the sum of the initial C, N, and O abundances as inferred from an EHe's
Fe abundance and standard relations for C, N, and O dependences on
initial Fe abundance for normal (i.e., H-normal) dwarfs. This is shown in
Figure 9. 

Oxygen: Oxygen abundances show a large spread: for example,
[O/H] at [Fe/H] $\sim$ 0 runs from about $+1$ to $-1$ and is, therefore,
generally at odds with a simple extrapolation from the N abundances that O
should be greatly depleted (Figure 10). 
The spread persists to lower [Fe/H] with
EHe stars with [O/H] $\simeq 0$ found at [Fe/H] $\simeq -2$. 

Neon: Our NLTE analysis shows Ne abundances are approximately  independent of a
star's Fe abundance (Figure 11). 
The spread in Ne abundance at a given Fe/H is
about  1 dex. Qualitatively, Ne is similar to O with respect to
spread and Fe-independence. Neon is not tightly correlated with the
O abundance but the O-richest stars include two of the most Ne-rich and,
perhaps significantly, are stars with a strong $s$-process enrichment
(V1920\,Cyg and LSE\,78).

Mg to Ni: Abundances of these metals (relative to Fe) follow the relations
determined from analyses of Galactic disk and
halo stars. In particular, the so-called $\alpha$-elements (Mg, Si, S, Ca,
and Ti) in EHe stars fall quite well on the established
[$\alpha$/Fe] versus [Fe/H] trends.

Iron: EHe stars span the metallicity range [Fe/H] $= -0.3$ to $-2.0$.

$s$-process: Several EHe stars appear enriched in $s$-process
products. V1920\,Cyg and LSE\,78, for example, show fifty-fold overabundances
of the lighter $s$-process products Y and Zr. 

Other intriguing abundance anomalies are provided from analyses
of RCB and HdC stars which would seem probable relatives of the
EHe stars. These anomalies which are undetectable in the EHe stars
because the spectroscopic signatures vanish at the
higher temperatures include:

$^{18}$O:  A spectacular anomaly is the extraordinarily high $^{18}$O
abundance seen in cool HdC and RCBs: $^{18}$O/$^{16}$O $\simeq 0.5$
in extreme cases \citep{clay2005,clay2007,garc2009,garc2010}.
Of course, the O isotopic abundance ratio requiring
detection of the CO molecule is not
measureable for either warm RCBs or the EHes. 

Lithium:  Similarly,
measurement of the Li abundance  demands a cool
atmosphere for detection of the Li\,{\sc i} resonance doublet
at 6707 \AA. Lithium is seen in one of the five HdC and in four of
approximately 30 known RCBs. 

Fluorine: A remarkable overabundance of F was discovered
by \citet{panF2006} for the cooler EHes and \citet{pan08} for
the warmer RCBs. The F\,{\sc i} lines vanish at the higher temperatures
of the hot EHes discussed here. The F abundances extend to
300 times the solar abundance and the maximum value
appears to be independent of
a star's Fe abundance. There is a star-to-star spread in F
abundances at a given [Fe/H]. 

Minority RCBs: A few RCBs show
highly anomalous [Si/Fe] and [S/Fe] ratios. Such stars were called
minority RCBs by \citet{lambrao94} with examples
including the RCB V CrA with [Si/Fe] $\simeq$ [S/Fe] $\simeq +2$  where
$\simeq +0.3$ is expected for normal H-rich stars of the same metallicity \citep{raolamb2008}.
None of the analyzed EHe stars has this minority characteristic but a larger
sample may uncover an example. The hot RCB DY\,Cen, a star known for its reluctance to
decline from maximum light, is a minority
star and might almost be called a EHe star.

\subsection{The Double-degenerate scenario}

\subsubsection{A recipe}

In the simplest implementation of the DD scenario, a He white dwarf (WD)
merges with a C-O WD to produce a EHe star without 
nucleosynthesis occurring during the merger, an occurrence which we term a cold
merger. Subsequent evolution of the H-deficient supergiant star is assumed not to result in
further changes of surface composition. Under these assumptions, it is possible to 
predict the composition of the EHe star created by the merger. 
The merger seems certain to result in a rapidly rotating compact
object. Expansion of the envelope to produce the EHe giant
will through conservation of angular momentum result in a less rapidly
rotating star. Here, one recalls that slowly rotating normal giants evolve from 
parts of the main sequence where rapidly rotating stars are common.
Line profiles of the EHes indicate a high macroturbulent velocity 
to which the projected rotational velocity may be a significant
contributor, e.g., BD$-9^\circ$4395's line profiles suggest the projected
rotational velocity is about 40 km s$^{-1}$.

In this picture, the merger involves mixing without
nuclear cooking of two principal ingredients: the He WD  and the
former He-shell of the C-O WD.\footnote{Here, the He WD has a He-core and is to be
distinguished from a DB WD, a star with a He atmosphere but a C-O core.}  
A third potential  ingredient may be provided
by those layers of the C-O WD immediately below its He-shell which may be
disturbed and mixed with the accreted He WD during the merger. One or both WDs
before the merger may contain a thin outer layer of H-rich material
but since H is very deficient in the merged star, we may neglect
these H-rich layers in the following discussion of  abundances
of major elements resulting from a cold merger. 
Gravitational settling may occur in both the He and the C-O WDs ahead of
the merger. Effects of settling in the He WD will be erased when the 
star is merged with the C-O WD. The merger event presumably stirs up the
C-O WD's He shell and again the effects of gravitational settling are
negated. Our recipe assumes that the He WD is thoroughly mixed with the 
thin He-shell of the C-O WD.
Implications of mergers involving H-rich layers
are explored briefly in an attempt to account for the Li-rich RCB stars.  
To predict the merged star's composition we need the masses and
compositions of the 
ingredients.    

\subsubsection{One ingredient - The He WD}

In principle, a He WD is created from low
mass main sequence stars but this requires a time
exceeding the age of the Galaxy. Thus, the  He WD in a DD scenario
must be a product of a binary system experiencing 
mass loss and probably
mass transfer. \citet{iben97} predict the mass distribution
of He  and C-O WDs expected to  result from evolution of close binaries: $M$(He) $\simeq
0.3\pm0.1 M_\odot$ and
$M$(C-O)  $\simeq 0.6\pm0.1 M_\odot$.
 Such a He  WD's composition is dominated by He and N:
the mass fraction  $\mu$(He) of He is essentially unity and, thanks to H-burning by
the CNO-cycles, the  N abundance is
the sum of the initial C, N, and O abundances,
say, $\mu$(CNO)$_0$ where $0$ denotes that the initial C, N, and O
abundances will be dependent on the initial metallicity (here inferred
from the Fe abundance). Mass fractions of C and O in the He white dwarf
 may be taken to be zero.
Heavier elements will have their initial mass
fractions. Thus, the mass of helium and nitrogen contributed to the
merged star assuming a conservative cold merger are essentially   $M$(He) and 
$\mu$(CNO)$_0$$M$(He), respectively.

\subsubsection{Another ingredient - The He shell of the C-O WD}

For the He shell of the C-O WD,   
estimates of
the mass and composition of the He shell  should
be obtained from calculations of binary star evolution that result in
appropriate  He and C-O WD pairs, but understandably
 such calculations appear not to have
been reported. Therefore, we take estimates from calculations for 
the inner regions of single stars in their AGB phase prior to loss of their
H-rich envelopes. In such cases, 
the mass of the He shell is approximately 0.02$M_\odot$ for 
1-3 $M_\odot$ stars but decreases to 0.002$M_\odot$ for the more massive
stars. We denote this mass by $M$(C-O:He). 
 
Early calculations showed that the He shell was primarily comprised
of $^4$He and $^{12}$C with mass fractions of about 0.75 and 0.20,
respectively, with $^{16}$O having a mass fraction of only about
0.01 \citep{sch79}. We make use of calculations  by
\citet{karakas2010a} (and private communication) for stars with masses
of 1 to 6 $M_\odot$ and with initial compositions $Z$=0.0001 to
0.02. Adopted compositions are the average of the He-shell's composition 
just prior to third dredge-up and at the point at which the star leaves
the AGB.  Mass fractions of $^4$He, $^{12}$C and $^{16}$O are
consistent with Sch\"{o}nberner's estimates.      
In Karakas et al.'s calculations,
the He mass fraction
$\mu$(He)$_{\rm {C-O:He}}$ is about 0.75, almost independent of mass
and composition. The $^{12}$C mass fraction $\mu$(C)$_{\rm {C-O:He}}$
$\simeq$ 0.20, again with little dependence on mass and composition.
$^{14}$N is effectively cleansed from the region and we take its mass fraction to
be zero.
The
$^{16}$O mass fraction is  $\mu$(O)$_{\rm {C-O:He}}$ $\simeq 0.005$.

Of particular interest to attempts
to match the EHe compositions is that the He shell has 
enhanced $^{22}$Ne and $^{19}$F abundances. 
Mass fractions of these two nuclides are dependent on the mass of the
initial star but are not particularly sensitive to the initial metal
mass fraction $Z$.
The $^{22}$Ne is synthesised from $^{14}$N by $\alpha$-capture, first
to $^{18}$O and then to $^{22}$Ne. Its abundance peaks  in stars of about
3$M_\odot$ reaching a mass fraction of about 0.05, a value not
greatly dependent on the initial metallicity of the star. The $^{22}$Ne
mass fraction decreases to lower initial stellar masses by a factor
that is metallicity dependent: at $Z=0.008$, the mass fraction for
a 1$M_\odot$ star is a factor of six below that for a 3$M_\odot$ star.
At higher masses than 3$M_\odot$, $^{22}$Ne is destroyed by $\alpha$-particles
and converted to $^{25}$Mg and $^{26}$Mg. 

Synthesis of $^{19}$F occurs from $^{15}$N by $^{15}$N($\alpha,\gamma$)$^{19}$F
in competition with $^{19}$F($\alpha$,p)$^{22}$Ne with $^{15}$N produced
by either  
$^{14}$N(n,p)$^{14}$C($\alpha,\gamma$)$^{18}$O(p,$\alpha$)$^{15}$N or
$^{14}$N($\alpha,\gamma$)$^{18}$F($\beta^+$)$^{18}$O(p,$\alpha$)$^{15}$N with
neutrons from $^{13}$C($\alpha$,n)$^{16}$O and protons from 
$^{14}$N(n,p)$^{14}$C.
The $^{19}$F mass fraction is a maximum for $M\simeq 3M_\odot$
decreasing by about  factors of 20-30 for 1$M_\odot$ and  10 for
6$M_\odot$. At maximum, the mass fraction 
$\mu$(F)$_{\rm {C-O:He}}$ $\simeq$ 1$\times$10$^{-4}$. 

Other series of calculations for AGB stars introduce
convective overshoot at the base of the He-burning thermal
pulse in the He shell.\footnote{This episode of convective overshoot
is to be distinguished from convective overshoot at the base
of the H-rich convective envelope into the top of the He shell
of an  AGB star. This affects operation of the $s$-process in the
He shell between thermal pulses and also the composition of the
surface layers - see \citet{karakas2010b} for a discussion.}
Overshoot necessarily brings more $^{16}$O
(and $^{12}$C) into the shell from the top of the C-O
core. Various implementations of convective overshoot have
been reported in the literature: for example, \citet{herwig2000}
(see also \citet{herwig2006}) reports for a  star of initial mass 3$M_\odot$
that the He shell at the last thermal pulse has mass fractions
of 0.41 and 0.18 for $^{12}$C and $^{16}$O, respectively, showing
an order of magnitude increase in the $^{16}$O mass fraction
from the calculation without this convective overshoot. One may
anticipate the possibility that O abundances in EHe stars may
offer an indirect test of calculations with and without convective
overshoot extending into the C-O core.

\subsubsection{Mixing the ingredients}

With estimates of the compositions of the two principal ingredients,
we may predict the
outcomes of a cold merger. As noted above, we may set H aside
because its abundance in the EHe star can be readily accounted for
by adding a small H-rich skin to the He WD and/or the C-O WD. 
Estimates for other elements are as follows:

The C/He ratio:  Helium is provided overwhelmingly by the
He WD and C exclusively by the C-O WD's He shell. The predicted C/He ratio by
number is 

\begin{equation}
\frac{\rm C}{\rm He} \simeq  \frac {A(\rm He)} {A(\rm C)} \frac{\mu(\rm C)_{\rm {C-O:He}}{\it M}(\rm {C-O:He})}
    {\it M(\rm He)}
\end{equation}

where $A$(X) denotes the atomic weight of X. 

With $\mu(\rm C)_{\rm (C-O:He)}\simeq 0.2$, $M{\rm {C-O:He}} \simeq 0.02 M_\odot$, and
$M(\rm He) \simeq 0.3M_\odot$, C/He $\simeq 0.4$\%.
Since the observed range of the C/He
ratio is from 0.3\% to 1.0\%, our prediction
accounts well for the lower end of the observed range. It is
possible to account for the upper end with not implausibly different choices
for the three variables.
If additional $^{12}$C is needed, it may be provided by  the third
ingredient, the `surface' layers of the C-O white dwarf where the $^{12}$C
mass fraction may approach 0.5 to 0.8: For example, an additional contribution
of 0.01$M_\odot$ with a $^{12}$C mass fraction of 0.5 raises the
C/He to 1\% after the merger.

Although not present in our sample of EHes for which we have derived the Ne
abundance, two EHes -- V652\,Her and HD\,144941 - as noted above have lower
C/He ratios  by two orders of magnitude: $C/He \simeq 0.003$\%. An interpretation
is that these stars result from a DD scenario involving a pair of He white
dwarfs. An alternative possibility based on the above equation is that the C-O
white dwarf in the merger had an unusually small He shell around the C-O white
dwarf, as might be anticipated for a star of intermediate mass.  

The N abundance: Nitrogen, a product of CNO-cycling,
 is contributed by the He WD which also
is the dominant contributor of mass to the merger and, hence, the
leading supplier of a reference element such as Fe. These circumstances
explain quite naturally why the  observed N abundance  in stars of
different Fe abundance is equal to the sum of the initial
C, N, and O abundances.

The oxygen abundance: The O abundance is given by

\begin{equation}
\frac{\rm O}{\rm He} \simeq  \frac {A(\rm He)} {A( \rm O)} \frac{\mu(\rm O)_{\rm C-O:He}{\it M}(\rm {C-O:He})}
    {\it M(\rm He)}
\end{equation}

With $\mu(\rm O)_{\rm (C-O:He)}\simeq 0.005$, $M{\rm (C-O:He)} \simeq 0.02 M_\odot$, and
$M(\rm He) \simeq 0.3M_\odot$, O/He $\simeq 0.008$\%
corresponding to an abundance $\log\epsilon$(O) $\simeq$ 7.5, a value at the
lower bound of the observed abundances. 

There appear to be two possibilities by which to increase the predicted
O abundances: (i) add the third ingredient, i.e., surface layers from the
C-O WD, when, for example, a mass of 0.01$M_\odot$ and an $^{16}$O
 mass fraction of
0.5 raises the O abundance to 9.2, the maximum observed value;
(ii) increase  the O mass fraction in the He shell to 0.2, a value
expected if convective overshoot during the AGB phase extends into the C-O core,  and
then the predicted O abundance is 8.1.
Both of these  possibilities may depend on individual
properties of the stars involved in the merger  and, thus,
might account for the observed spread in O abundances among
EHe (and RCB) stars. As long as the responsible agent is not
metallicity dependent, the lack of a trend of O abundance with
metallicity is accounted for.

The Neon abundance:  The 
Ne abundance is

\begin{equation}
\frac{\rm Ne}{\rm He} \simeq  \frac {A(\rm He)} {A(^{22}{\rm Ne})} \frac{\mu(\rm Ne)_{\rm C-O:He}{\it M}(\rm {C-O:He})+ \mu(\rm Ne)_0{\it M}(\rm He)}
    {\it M(\rm He)}
\end{equation}

where contributions from the He-shell of the C-O white dwarf and the
original Ne content of the He white dwarf are included.

With $\mu(\rm Ne)_{\rm (C-O:He)} \simeq 0.05$, $M({\rm C-O:He}) \simeq 0.02M_\odot $ and $M(\rm He) \simeq 0.3M_\odot$, 
Ne/He $\simeq 0.06\%$
corresponding to an abundance $\log\epsilon$(Ne) $\simeq$ 8.4, a value 
equal to the upper bound of the observed abundances (Table 8, and Figure 11). Lower
Ne abundances are readily achieved because the Ne mass fraction is
lower in all but 3$M_\odot$ stars. Karakas's predictions 
encompass an order of magnitude range in Ne mass fractions
and the  observed Ne abundances from our (small) sample of EHe stars
show a factor of five spread. In all cases but two (HD\,124448 and PV\,Tel),
the Ne
abundance is appreciably greater than the presumed initial abundance
based on the Fe abundance and likely relation between initial Ne and Fe
abundances.

The Fluorine abundance: Fluorine is not detectable in these
hot EHes but its abundance is known for cooler EHes and
the warmer RCBs, as noted above. The maximum mass fraction for F in the
He shell gives an abundance $\log\epsilon$(F) $\simeq$  5.7 but observed
abundances are about 1 dex higher. This may suggest that the
nucleosynthesis of F has been underestimated in the He shell, or
the derived F abundance for cooler EHes has been overestimated. 
This gap might also be bridged
by NLTE calculations of F\,{\sc i} line formation; such calculations for
Ne\,{\sc i}, an atom with a not dissimilar atomic structure, show an
almost one dex reduction of the observed LTE abundances.

The $^{18}$O abundance:  Extraordinary amounts of $^{18}$O are present  with
the $^{18}$O abundance 
exceeding that of $^{16}$O in several stars. 
The $^{18}$O/$^{16}$O ratio is higher and the $^{18}$O
abundance lower in most of the RCBs where CO lines are detectable; 
\citet{garc2010} speculate that a late dredge-up in
a HdC  star is responsible for these differences between HdC and RCB stars.
These high $^{18}$O abundances cannot be explained by the two or even
the three ingredient recipe describing the DD scenario as a cold merger.
One may wonder if the C-O WD's He shell has a composition that is not
a complete replica of a He shell of a single star, the model we
have adopted for these calculations. Perhaps, the synthesis of $^{22}$Ne
from $^{14}$N is incomplete and some $^{18}$O remains. But one wonders if
this suspicion may be reconciled with the fact that
 the maximum $^{18}$O abundances for the HdC
stars ($\log\epsilon(^{18}$O) $\simeq 8.6$) are greater than the Ne abundances 
($\log\epsilon$(Ne)$ \simeq 8.5$) in the EHe stars. Is such fine tuning
possible? 

\citet{clay2005} considered $^{18}$O synthesis to occur by nuclear
processing during  accretion of He WD material
by the C-O WD. In the
processing, $^{14}$N is converted to $^{18}$O by $\alpha$-capture.
The observed $^{18}$O abundances are not sufficiently
great to have sensibly reduced the $^{14}$N abundances: the $^{18}$O
abundances are factors of 4 to 20 less than the $^{14}$N abundances.
However, fine-tuning is required in order not to deplete entirely the $^{14}$N
supply and also to prevent  conversion of significant amounts of
 $^{18}$O by $\alpha$-capture
to $^{22}$Ne. 

The $s$-process: Enrichment of $s$-process nuclides  likely in the
C-O WD's He shell  will be diluted by the absence of enrichment in the
He WD. Thus, the enrichment will be diluted by about an order or
magnitude if $M$(He)$\simeq 0.3M_\odot$ and $M$(He)$_{C-O} \simeq 0.02M_\odot$. 
Two of the EHes are observed enriched in Y and Zr  50-fold and another
10-fold. Other stars have lower enrichment levels. Although a 100- to
500-fold enrichment is within expected levels for isolated AGB
stars, the general
lack of a large $s$-process
enrichment of these EHes, as well as the cool EHes \citep{pan2001}
and the warmer RCBs \citep{asp00} implies that the He shell of the
C-O WD was not itself greatly $s$-process enriched. This would seem
to confirm suspicions that the He shells participating in the
DD scenario may not  be near-copies of He shells of isolated AGB stars.

Lithium: Lithium is, of course, not observable in EHe stars but its presence in
several RCBs and in one of five known HdCs suggests that it is
probably present in at least some EHe stars. Therefore, the challenge 
exists to account for the Li abundance in the DD scenario. 

An initial supposition is that the Li was present in the H-rich skin
around the white dwarfs. Then, the observed Li/H ratio for the
RCB star is the mass-weighted mean of the Li/H ratio in the two H-rich
skins. For the four RCBs with Li, $\log$\,Li/H ranges from $-1.7$ for RZ\,Nor to
$-4.8$ for SU\,Tau \citep{asp00} or Li abundances of 10.3 to 7.2 on the
usual logarithmic scale where the H abundance is 12.0. 
Such
extraordinarily high Li abundances are observed nowhere else: for example,
the Li-rich carbon stars have Li abundances {\it only} (!)
 in the range 3 to 5  \citep{abia99}.

In the usual scheme of Li
synthesis known as the Cameron-Fowler mechanism \citep{cam1971}, Li as $^7$Li is
synthesised from $^3$He by the chain 
$^3$He($^4$He,$\gamma)^7$Be($e^-,\nu)^7$Li. 
The potential reservoir of $^3$He is the star's
original supply of $^3$He and $^2$H (which is burnt to $^3$He in 
pre-main sequence phase) and
additional $^3$He provided by operation of the $pp$-chains in low mass
main sequence stars. 
The original $^3$He/H after $^2$H burning will 
have been about 2$\times 10^{-5}$.
In low mass stars, a layer of enriched $^3$He exists outside the
H-burning core. The abundance can reach 10$^{-3}$ of that of H over a
shell about 0.2$M_\odot$ in thickness \citep{iben67}. 
Protons are not directly involved in the Cameron-Fowler mechanism but the
temperature of the synthesis site may be influenced by the mass exterior to
the site.

In the case of the Li-rich normal carbon stars, the site is the high temperature
base of the H-rich convective envelope in an intermediate-mass
AGB star. In this environment, the
observed range of Li abundances can be achieved. To achieve a higher
abundance (i.e., a higher Li/H ratio), it seems necessary to
reduce the mass of the H into which products ($^7$Be and then $^7$Li)
of $^3$He consumption
are mixed. Efficiency of $^7$Li production might be maximised were the
$^3$He consumed in a layer completely devoid of H; this would
remove the loss of $^7$Be and $^7$Li by proton capture but these nuclides
would still be prone to destruction by $\alpha$-capture. 
Further exploration of $^7$Li synthesis will require very detailed
calculations of nucleosynthesis in close binary systems.

\subsection{The Final-flash scenario}

Several classes of H-deficient hot luminous post-AGB stars are believed to
have resulted from a final-flash scenario. Compositions of these stars
offer a direct point of comparison for the EHe (also RCB and HdC) stars
and, therefore, a test of the final-flash scenario as an origin for
some EHe stars.

A valuable review of the observed compositions and theoretical origins
of hot H-deficient post-AGB stars was provided by \citet{werner2006}.
Their Table 1 clearly shows that the several families ([WCL], [WCE],
[WC]-PG1159, and PG1159) of such post-AGB stars  have compositions
differing in several distinctive ways from the compositions
of EHe and RCB stars. In particular, the relative C/He ratio is quite
different. The (He,C) mass fractions are roughly in the range
(0.30,0.60) to (0.85,0.15) whereas the EHes are close to
(0.98,0.02) \citep{werner2008}.  This contrast is, of course, very largely
attributable to the He contribution by the He white dwarf to the DD scenario.
  Several of the post-AGB stars show
a residue of their original H with a mass fraction of as much as
0.35, but a lower value is more common. (These are hot stars and, therefore,
detection of H lines is difficult.) In other respects, the compositions
of the post-AGB and EHe stars are more similar: the O mass fractions show a
star-to-star spread but the highest value for a post-AGB star (0.20,
\citet{werner2008}) is several times higher than the maximum value for a
EHe. The F and Ne abundances are similar for the two groups of stars.

Theoretical final flash models are discussed by \citet{werner2006}.
These models differ as to when the He-shell flash 
(thermal pulse) occurs that
restores the star to the post-AGB track as a EHe-like star.  If the
thermal pulse occurs when the star is on the AGB, it is termed a AFTP
and may account for the relatively H-rich stars known as hybrid-PG1159
stars. If the thermal pulse occurs in the post-AGB period of approximately
constant luminosity, it is
a LTP (L=Late) and is predicted to create a H-deficient star with
some H, say a mass fraction of about 0.02. FG\,Sge may have
experienced a LTP - recently! Finally, if the thermal
pulse is delayed until the star is on the WD cooling track, it is
a VLTP (V=very). Sakurai's object is  considered to have
undergone a VLTP - very recently! Although these post-AGB final flash
scenarios fail to account for the EHes, one may note some
similarities with the ideas behind the DD scenario. In particular,
convective extramixing into the `surface' layers of the C-O core
is invoked in both cases to account for the spread in the O
abundances, and high F and Ne abundances in the He shell around the
C-O core of the AGB star {\bf to} explain the F and Ne overabundances in
both kinds of H-deficient stars.

\section{Concluding remarks}

Understanding the origins of peculiar stars often awaits recognition of
predecessors and descendants along the evolutionary sequence. The sequence
may provide
clues not always provided by even detailed observational studies of a
single class of peculiar star in the sequence. Even more important than
expanded observational studies is the
development of theoretical understanding of relevant stellar models and
associated nucleosynthesis. The EHe stars illustrate well these
remarks.

The first EHe HD\,124448 was discovered by \citet{pop1942} at the
McDonald Observatory and this was followed ten years later by
discovery of  PV\,Tel, alias
HD\,168476,  by \citet{thack1952} at the Radcliffe Observatory. Today,
the register of EHes totals 
about 30 \citep{jeff96}.\footnote{H-deficient  binaries such as KS\,Per and
$\upsilon$ Sgr are an unrelated and even rarer type of star.}
Today, the
chemical compositions of EHe stars are quite well determined with our
estimates of Ne abundances adding one more data point.
Similarities in composition argue for an evolutionary link between
EHe and RCB stars and less convincingly between these stars and the
HdC stars of which only five are known and their spectra blessed with
a rich array of molecular lines render accurate abundance analysis difficult.
Theoretical ideas have centered on two possibilities: the DD- and the FF-scenarios.

While the FF scenario may account for H-deficient stars like FG\,Sge and
Sakurai's object (as noted above), the argument eliminating it as
the origin of EHe stars is a fusion of two principal points. First,
the compositions of H-deficient central stars of planetary
nebulae differ in critical aspects from those of EHe stars. As noted
in Section 5.3, the He and C mass fractions of the central stars differ
greatly from these quantities as found for EHe (and RCB) stars.
Second, the observed compositions of the central stars are reasonably
well accounted for by models of final He-shell flashes in post-AGB
stars. Thus, a safe conclusion  would appear to be: the FF scenario
cannot account for the EHe stars.

Identification of the EHe (and RCB) stars with the DD scenario
depends on the  correspondence between the
measured chemical compositions and semi-quantitative theoretical
estimates resulting from the merger of a He WD with a C-O WD. Predictions
for a cold merger seem especially sensitive to the adopted composition
of the He-shell around the C-O WD before the merger and the extent
to which mixing during the merger may incorporate material
from the surface layers of the C-O WD. An additional uncertainty concerns
the extent of nucleosynthesis occurring during the merger; the
proposal that the $^{18}$O seen in cool HdC and RCB stars is
synthesised from $^{14}$N during the merger was noted \citep{clay2007}.

Although work remains for quantitative spectroscopists to do, we close
with the thought that the larger challenges remain in the area of
theoretical modelling of single stars in order to refine prediction
for the several forms of the FF scenario and of double stars that
through common envelope stages and mass loss provide the necessary stage for the
DD scenario of a
 close binary of a He and a C-O WD
that merge with possibly a concluding episode of
nucleosynthesis to provide a EHe, RCB, or a HdC.

\acknowledgments
We thank the referee for a constructive report.
We are indebted to Amanda Karakas for providing details of her calculations of
AGB stars. We thank Katia Cunha for providing her Ne\,{\sc i} line list for
HD 35299.
GP thanks Baba Verghese for the help provided in installing the TLUSTY/SYNSPEC codes.
GP is thankful to Carlos Allende Prieto for his useful suggestions in making the
TLUSTY models. We thank Ivan Hubeny for testing the convergence of one of 
our NLTE TLUSTY model, and for all the discussions with him. GP is also
thankful to Simon Jeffery for making available Armagh LTE and other related
codes. Travel support for GP to visit UT Austin
and McDonald Observatory where a part of this work was carried out was provided by the
Isabel McCutcheon Harte Centennial Chair.
DLL thanks the Robert A. Welch Foundation for support through grant F-634.

\begin{deluxetable}{lccrcc}
\tabletypesize{\scriptsize}
\tablewidth{0pt}
\tablecolumns{6}
%\tablewidth{0pc}
\setcounter{table} {0}
\tablecaption{Photospheric line by line NLTE and LTE abundances, and
the line's measured equivalent width ($W_{\lambda}$) in m\AA\, for BD$+10^\circ 2179$}
\tablehead{
\colhead{} & \colhead{$\chi$} & \colhead{} & \colhead{$W_{\lambda}$} &
\multicolumn{2}{c}{log $\epsilon(\rm X)$}\\
\cline{5-6} \\
\colhead{Line} & \colhead{(eV)} & \colhead{log $gf$} & \colhead{(m\AA)} & \colhead{NLTE\tablenotemark{a}} &
\colhead{LTE\tablenotemark{b}}}
\startdata
H\,{\sc i} $\lambda 3970.072$ & 10.199 & $-$0.993 &  90 & 8.26 &  8.60 \\
H\,{\sc i} $\lambda 4101.734$ & 10.199 & $-$0.753 & 146 & 8.28 &  8.60 \\
H\,{\sc i} $\lambda 4340.462$ & 10.199 & $-$0.447 & 242 & 8.32 &  8.64 \\
H\,{\sc i} $\lambda 4861.323$ & 10.199 & $-$0.020 & 430 & 8.54 &  8.88 \\
Mean... & \nodata & \nodata & \nodata & 8.35$\pm$0.13   & 8.68$\pm$0.13 \\
C\,{\sc i} $\lambda 4932.049$ & 7.685 & $-$1.658 & 13 & 9.33 &  9.41 \\
C\,{\sc i} $\lambda 5052.167$ & 7.685 & $-$1.303 & 28 & 9.37 &  9.42 \\
Mean... & \nodata & \nodata & \nodata & 9.35$\pm$0.03 & 9.42$\pm$0.01 \\
C\,{\sc ii} $\lambda 3918.980$ & 16.333 & $-$0.533 & 286 & 9.08 &  9.30 \\
C\,{\sc ii} $\lambda 3920.690$ & 16.334 & $-$0.232 & 328 & 9.01 &  9.28 \\
C\,{\sc ii} $\lambda 4017.272$ & 22.899 & $-$1.031 & 43  & 9.40 &  9.40 \\
C\,{\sc ii} $\lambda 4021.166$ & 22.899 & $-$1.333 & 27  & 9.40 &  9.40 \\
C\,{\sc ii} $\lambda 4306.330$ & 21.150 & $-$1.684 & 46  & 9.25 &  9.32 \\
C\,{\sc ii} $\lambda 4307.581$ & 20.150 & $-$1.383 & 77  & 9.33 &  9.42 \\
C\,{\sc ii} $\lambda 4313.100$ & 23.120 & $-$0.373 & 83  & 9.40 &  9.38 \\
C\,{\sc ii} $\lambda 4317.260$ & 23.120 & $-$0.005 & 113 & 9.35 &  9.36 \\
C\,{\sc ii} $\lambda 4318.600$ & 23.120 & $-$0.407 & 80  & 9.40 &  9.38 \\
C\,{\sc ii} $\lambda 4321.650$ & 23.120 & $-$0.901 & 45  & 9.43 &  9.41 \\
C\,{\sc ii} $\lambda 4323.100$ & 23.120 & $-$1.105 & 45  & 9.64 &  9.62 \\
C\,{\sc ii} $\lambda 4637.630$ & 21.150 & $-$1.229 & 75  & 9.21 &  9.27 \\
C\,{\sc ii} $\lambda 4867.066$ & 19.495 & $-$1.781 & 35  & 9.27 &  9.10 \\
C\,{\sc ii} $\lambda 5032.128$ & 20.922 & $-$0.143 & 174 & 9.40 &  9.45 \\
C\,{\sc ii} $\lambda 5035.943$ & 20.920 & $-$0.399 & 113 & 9.09 &  9.14 \\
C\,{\sc ii} $\lambda 5125.208$ & 20.150 & $-$1.597 & 51  & 9.36 &  9.40 \\
C\,{\sc ii} $\lambda 5126.963$ & 20.150 & $-$1.899 & 32  & 9.36 &  9.40 \\
C\,{\sc ii} $\lambda 5137.257$ & 20.701 & $-$0.911 & 91  & 9.32 &  9.36 \\
C\,{\sc ii} $\lambda 5139.174$ & 20.704 & $-$0.707 & 118 & 9.38 &  9.43 \\
C\,{\sc ii} $\lambda 5143.495$ & 20.704 & $-$0.212 & 166 & 9.31 &  9.38 \\
C\,{\sc ii} $\lambda 5145.165$ & 20.710 & $+$0.189 & 207 & 9.34 &  9.31 \\
C\,{\sc ii} $\lambda 5151.085$ & 20.710 & $-$0.179 & 169 & 9.37 &  9.37 \\
Mean... & \nodata & \nodata & \nodata & 9.32$\pm$0.14   & 9.36$\pm$0.11 \\
C\,{\sc iii} $\lambda 4186.900$ & 40.010 &$+$0.918 & 19  & 9.29 &  9.99 \\
C\,{\sc iii} $\lambda 4647.420$ & 29.535 &$+$0.070 & 51 &10.01? &  9.39 \\
C\,{\sc iii} $\lambda 4650.250$ & 29.535 &$-$0.151 & 42 &10.05? &  9.40 \\
C\,{\sc iii} $\lambda 4651.470$ & 29.535 &$-$0.629 & 27 &10.15? &  9.43 \\
Mean... & \nodata & \nodata & \nodata & 9.29$\pm$0.00   & 9.41$\pm$0.02 \\
N\,{\sc ii} $\lambda 3842.180$ & 21.150 & $-$0.692 & 30 &  8.21  &  8.13 \\
N\,{\sc ii} $\lambda 3955.851$ & 18.466 & $-$0.813 & 68 &  8.14  &  8.15 \\
N\,{\sc ii} $\lambda 3994.996$ & 18.498 & $+$0.208  & 139 & 7.87  &  7.95 \\
N\,{\sc ii} $\lambda 4179.670$ & 23.250 & $-$0.204 & 23  & 8.29  &  8.12 \\
N\,{\sc ii} $\lambda 4227.740$ & 21.600 & $-$0.061 & 42  & 8.07  &  7.95 \\
N\,{\sc ii} $\lambda 4447.030$ & 20.411 & $+$0.228  & 75  & 7.94  &  7.87 \\
N\,{\sc ii} $\lambda 4507.560$ & 20.666 & $-$0.817 & 25  & 8.24  &  8.12 \\
N\,{\sc ii} $\lambda 4601.480$ & 18.468 & $-$0.428 & 79  & 8.00  &  8.00 \\
N\,{\sc ii} $\lambda 4607.160$ & 18.464 & $-$0.507 & 73  & 8.01  &  8.01 \\
N\,{\sc ii} $\lambda 4613.870$ & 18.468 & $-$0.665 & 61  & 8.02  &  8.00 \\
N\,{\sc ii} $\lambda 4643.090$ & 18.484 & $-$0.359 & 88  & 8.05  &  8.06 \\
N\,{\sc ii} $\lambda 4654.531$ & 18.497 & $-$1.404 & 20  & 8.04  &  7.99 \\
N\,{\sc ii} $\lambda 4667.208$ & 18.497 & $-$1.533 & 20  & 8.17  &  8.12 \\
N\,{\sc ii} $\lambda 4674.908$ & 18.497 & $-$1.463 & 19  & 8.07  &  8.02 \\
N\,{\sc ii} $\lambda 4779.720$ & 20.650 & $-$0.587 & 34  & 8.27  &  8.14 \\
N\,{\sc ii} $\lambda 4781.190$ & 20.650 & $-$1.308 &  9  & 8.22  &  8.10 \\
N\,{\sc ii} $\lambda 4788.130$ & 20.650 & $-$0.363 & 35  & 8.07  &  7.94 \\
N\,{\sc ii} $\lambda 4810.310$ & 20.660 & $-$1.084 & 17  & 8.35  &  8.21 \\
N\,{\sc ii} $\lambda 4895.117$ & 17.877 & $-$1.338 & 18  & 7.76  &  7.72 \\
N\,{\sc ii} $\lambda 5002.700$ & 18.480 & $-$1.022 & 42  & 8.16  &  8.09 \\
N\,{\sc ii} $\lambda 5007.328$ & 20.940 & $+$0.171  & 52  & 7.98  &  7.84 \\
N\,{\sc ii} $\lambda 5010.620$ & 18.470 & $-$0.607 & 71  & 8.12  &  8.08 \\
N\,{\sc ii} $\lambda 5025.659$ & 20.666 & $-$0.547 & 28  & 8.17  &  8.03 \\
N\,{\sc ii} $\lambda 5045.090$ & 18.460 & $-$0.407 & 87  & 8.10  &  8.07 \\
Mean... & \nodata & \nodata & \nodata & 8.10$\pm$0.14 & 8.03$\pm$0.11 \\
O\,{\sc ii} $\lambda 4072.157$ & 25.643 & $+$0.552  & 26  & 7.92  &  7.60 \\
O\,{\sc ii} $\lambda 4185.449$ & 28.351 & $+$0.604  & 11  & 8.14  &  7.69 \\
O\,{\sc ii} $\lambda 4189.789$ & 28.354 & $+$0.717  & 20  & 8.47  &  7.99 \\
O\,{\sc ii} $\lambda 4336.860$ & 22.973 & $-$0.763 & 11  & 7.93  &  7.69 \\
O\,{\sc ii} $\lambda 4345.567$ & 22.979 & $-$0.346 & 18  & 7.81  &  7.59 \\
O\,{\sc ii} $\lambda 4349.426$ & 22.993 & $+$0.060  & 32  & 7.79  &  7.61 \\
O\,{\sc ii} $\lambda 4366.888$ & 22.993 & $-$0.348 & 15  & 7.71  &  7.49 \\
O\,{\sc ii} $\lambda 4414.901$ & 23.435 & $+$0.172  & 33  & 7.88  &  7.66 \\
O\,{\sc ii} $\lambda 4416.973$ & 23.413 & $-$0.077 & 26  & 7.97  &  7.72 \\
O\,{\sc ii} $\lambda 4641.817$ & 22.973 & $+$0.055  & 32  & 7.87  &  7.69 \\
O\,{\sc ii} $\lambda 4649.143$ & 22.993 & $+$0.308  & 41  & 7.79  &  7.65 \\
Mean... & \nodata & \nodata & \nodata & 7.93$\pm$0.21  & 7.67$\pm$0.12 \\
Ne\,{\sc i} $\lambda 5852.488$ & 16.850 & $-$0.490 & 61  & 7.78  &  8.54 \\
Ne\,{\sc i} $\lambda 6074.338$ & 16.670 & $-$0.500 & 66  & 7.87  &  8.56 \\
Ne\,{\sc i} $\lambda 6143.063$ & 16.620 & $-$0.100 & 115 & 7.84  &  8.64 \\
Ne\,{\sc i} $\lambda 6163.594$ & 16.710 & $-$0.620 & 65  & 7.99  &  8.69 \\
Ne\,{\sc i} $\lambda 6266.495$ & 16.710 & $-$0.370 & 97  & 7.98  &  8.77 \\
Ne\,{\sc i} $\lambda 6334.428$ & 16.620 & $-$0.320 & 90  & 7.91  &  8.63 \\
Ne\,{\sc i} $\lambda 6382.992$ & 16.670 & $-$0.240 & 103 & 7.89  &  8.70 \\
Ne\,{\sc i} $\lambda 6402.246$ & 16.620 & $+$0.330 & 163 & 7.68  &  8.73 \\
Ne\,{\sc i} $\lambda 6506.528$ & 16.670 & $-$0.030 & 121 & 7.81  &  8.68 \\
Ne\,{\sc i} $\lambda 7032.413$ & 16.620 & $-$0.260 & 116 & 7.95  &  8.86 \\
Mean... & \nodata & \nodata & \nodata & 7.87$\pm$0.10 & 8.68$\pm$0.10 \\
\enddata
\tablenotetext{a}{($T_{\rm eff}$, $\log g$, $\xi$)=(16375, 2.45, 7.5)}
\tablenotetext{b}{($T_{\rm eff}$, $\log g$, $\xi$)=(17000, 2.60, 7.5)}
\end{deluxetable}

\begin{deluxetable}{lccrcc}
\tabletypesize{\scriptsize}
\tablewidth{0pt}
\tablecolumns{6}
%\tablewidth{0pc}
\setcounter{table} {1}
\tablecaption{Photospheric line by line NLTE and LTE abundances, and
the line's measured equivalent width ($W_{\lambda}$) in m\AA\, for BD$-9^\circ 4395$}
\tablehead{
\colhead{} & \colhead{$\chi$} & \colhead{} & \colhead{$W_{\lambda}$} &
\multicolumn{2}{c}{log $\epsilon(\rm X)$}\\
\cline{5-6} \\
\colhead{Line} & \colhead{(eV)} & \colhead{log $gf$} & \colhead{(m\AA)} & \colhead{NLTE\tablenotemark{a}} &
\colhead{LTE\tablenotemark{b}}}
\startdata
H\,{\sc i} $\lambda 3970.072$ & 10.199 & $-$0.993 & 60 & 9.16 &   8.90 \\
H\,{\sc i} $\lambda 4101.734$ & 10.199 & $-$0.753 & 65 & 9.12 &   8.72 \\
H\,{\sc i} $\lambda 4861.323$ & 10.199 & $-$0.020 & 200 & 9.10 &   8.75 \\
Mean... & \nodata & \nodata & \nodata & 9.13$\pm$0.03 &  8.79$\pm$0.10 \\
C\,{\sc ii} $\lambda 4313.106$ & 23.116 & $-$0.373 & 108 & 9.27 &   9.43 \\
        &         &         &         &               &               \\
C\,{\sc ii} $\lambda 4325.832$ & 23.119 & $-$0.373 &    &       &      \\
C\,{\sc ii} $\lambda 4326.164$ & 23.116 & $-$0.407 & 131 & 9.04 &   9.17   \\
        &         &         &         &               &               \\
C\,{\sc ii} $\lambda 4374.281$ & 24.654 & $+$0.660\tablenotemark{c} &     &      &      \\
C\,{\sc ii} $\lambda 4375.008$ & 24.658 & $-$0.610\tablenotemark{c} & 189 & 9.01 &   9.06    \\
Mean... & \nodata & \nodata & \nodata & 9.11$\pm$0.14  & 9.22$\pm$0.19 \\
C\,{\sc iii} $\lambda 4515.352$ & 39.401 & $-$0.756 &     &      &      \\
C\,{\sc iii} $\lambda 4515.811$ & 39.402 & $-$0.279 & 28  & 8.92 &     8.71 \\
        &         &         &         &               &               \\
C\,{\sc iii} $\lambda 4647.418$ & 29.535 & $+$0.070 & 354 & 9.04 &  9.15      \\
C\,{\sc iii} $\lambda 4651.473$ & 29.535 & $-$0.629 & 219 & 9.23 &  9.12 \\
Mean... & \nodata & \nodata & \nodata & 9.06$\pm$0.16 &  8.99$\pm$0.25 \\
N\,{\sc ii} $\lambda 4601.478$ & 18.466 & $-$0.428 & 125  & 7.66 &  7.98     \\
N\,{\sc ii} $\lambda 4607.153$ & 18.462 & $-$0.507 & 103  & 7.60 &  7.93     \\
N\,{\sc ii} $\lambda 4621.393$ & 18.466 & $-$0.514 & 98   & 7.57 &  7.91     \\
N\,{\sc ii} $\lambda 4630.539$ & 18.483 & $+$0.094 & 224  & 7.70 &  7.98     \\
Mean... & \nodata & \nodata & \nodata & 7.63$\pm$0.06 &  7.95$\pm$0.04 \\
N\,{\sc iii} $\lambda 4103.390$ & 27.438 & $-$0.359 & 84  & 7.69 &  7.88     \\
N\,{\sc iii} $\lambda 4195.740$ & 36.842 & $-$0.004 & 20  & 7.94 &  8.39     \\
N\,{\sc iii} $\lambda 4200.070$ & 36.856 & $+$0.250 & 20  & 7.69 &  8.14     \\
Mean... & \nodata & \nodata & \nodata & 7.77$\pm$0.14  & 8.14$\pm$0.26 \\
O\,{\sc ii} $\lambda 4072.153$ & 25.650 & $+$0.552 & 134  & 8.13 &  7.78    \\
O\,{\sc ii} $\lambda 4078.842$ & 25.638 & $-$0.284 &  32  & 7.99 &  7.77    \\
O\,{\sc ii} $\lambda 4085.112$ & 25.650 & $-$0.189 &  46  & 8.10 &  7.86    \\
        &         &         &         &               &               \\
O\,{\sc ii} $\lambda 4104.724$ & 25.837 & $-$0.302 &      &      &      \\
O\,{\sc ii} $\lambda 4104.990$ & 25.837 & $-$0.015 &  75  & 8.05 &  7.80     \\
        &         &         &         &               &               \\
O\,{\sc ii} $\lambda 4303.833$ & 28.822 & $+$0.640\tablenotemark{c} &  40  & 7.97 &  7.67    \\
O\,{\sc ii} $\lambda 4331.857$ & 28.512 & $-$0.136 &  42  & 8.34 &  8.41    \\
O\,{\sc ii} $\lambda 4336.859$ & 22.979 & $-$0.763 &  76  & 8.08 &  8.13    \\
O\,{\sc ii} $\lambda 4342.009$ & 28.883 & $+$0.820\tablenotemark{c} &  44  & 7.84 &  7.51    \\
O\,{\sc ii} $\lambda 4345.560$ & 22.979 & $-$0.346 & 119  & 7.98 &  8.00    \\
O\,{\sc ii} $\lambda 4349.426$ & 22.999 & $+$0.060 & 212  & 8.09 &  8.10    \\
        &         &         &         &               &               \\
O\,{\sc ii} $\lambda 4351.457$ & 25.661 & $-$1.004 &      &      &       \\
O\,{\sc ii} $\lambda 4351.260$ & 25.661 & $+$0.227 &  92  & 8.13 &  7.83     \\
        &         &         &         &               &              \\
O\,{\sc ii} $\lambda 4366.895$ & 22.999 & $-$0.348 & 134  & 8.08 &  8.09   \\
O\,{\sc ii} $\lambda 4414.899$ & 23.442 & $+$0.172 & 206  & 7.86 &  8.06   \\
O\,{\sc ii} $\lambda 4416.975$ & 23.419 & $-$0.077 & 162  & 7.91 &  8.07   \\
O\,{\sc ii} $\lambda 4452.378$ & 23.442 & $-$0.788 &  56  & 7.92 &  8.09   \\
O\,{\sc ii} $\lambda 4590.974$ & 25.661 & $+$0.350 & 132  & 8.38 &  8.01   \\
        &         &         &         &               &              \\
O\,{\sc ii} $\lambda 4595.957$ & 25.661 & $-$1.033 &      &      &     \\
O\,{\sc ii} $\lambda 4596.177$ & 25.661 & $+$0.200 & 109  & 8.32 &  7.99   \\
        &         &         &         &               &              \\
O\,{\sc ii} $\lambda 4609.373$ & 29.069 & $+$0.670\tablenotemark{c} &  54  & 7.97 &  7.88   \\
O\,{\sc ii} $\lambda 4661.632$ & 22.979 & $-$0.278 & 141  & 8.01 &  8.07   \\
O\,{\sc ii} $\lambda 4705.346$ & 26.249 & $+$0.477 &  94  & 7.42 &  7.78   \\
O\,{\sc ii} $\lambda 4941.072$ & 26.554 & $-$0.053 &  45  & 8.27 &  7.97   \\
Mean... & \nodata & \nodata & \nodata & 8.04$\pm$0.21 & 7.95$\pm$0.20 \\
Ne\,{\sc i} $\lambda 5852.488$ & 16.850 & $-$0.490 & 21  & 8.20  &  8.65     \\
Ne\,{\sc i} $\lambda 6143.063$ & 16.620 & $-$0.100 & 63  & 8.18  &  8.77     \\
Ne\,{\sc i} $\lambda 6266.495$ & 16.710 & $-$0.370 & 42  & 8.11  &  8.85     \\
Ne\,{\sc i} $\lambda 6334.428$ & 16.620 & $-$0.320 & 36  & 8.13  &  8.71     \\
Ne\,{\sc i} $\lambda 6382.992$ & 16.670 & $-$0.240 & 48  & 8.17  &  8.79     \\
Ne\,{\sc i} $\lambda 6402.246$ & 16.620 & $+$0.330 & 143 & 8.19  &  8.89     \\
Ne\,{\sc i} $\lambda 6506.528$ & 16.670 & $-$0.030 & 65  & 8.13  &  8.75     \\
Ne\,{\sc i} $\lambda 6598.953$ & 16.850 & $-$0.360 & 34  & 8.39  &  8.79     \\
Ne\,{\sc i} $\lambda 7032.413$ & 16.620 & $-$0.260 & 46  & 8.16  &  8.80     \\
Mean... & \nodata & \nodata & \nodata & 8.18$\pm$0.08 &  8.78$\pm$0.07 \\
Ne\,{\sc ii} $\lambda 4379.552$ & 34.802 & $+$0.780 & 38  & 8.05  &   8.04   \\
Ne\,{\sc ii} $\lambda 4413.113$ & 34.833 & $+$0.520 & 22  & 8.02  &   8.01   \\
Mean... & \nodata & \nodata & \nodata & 8.04$\pm$0.02 & 8.03$\pm$0.02 \\
\enddata
\tablenotetext{a}{($T_{\rm eff}$, $\log g$, $\xi$)=(24300, 2.65, 17.5)}
\tablenotetext{b}{($T_{\rm eff}$, $\log g$, $\xi$)=(24800, 2.85, 23.0)}
\tablenotetext{c}{\citet{wiese66}}
\end{deluxetable}

\begin{deluxetable}{lccrcc}
\tabletypesize{\scriptsize}
\tablewidth{0pt}
\tablecolumns{7}
%\tablewidth{0pc}
\setcounter{table} {2}
\tablecaption{Photospheric line by line LTE neon abundances, 
NLTE neon abundance, and
the line's measured equivalent width ($W_{\lambda}$) in m\AA\, for LSE\,78}
\tablehead{
\colhead{} & \colhead{$\chi$} & \colhead{} & \colhead{$W_{\lambda}$} &
\multicolumn{2}{c}{log $\epsilon(\rm Ne)$}\\
\cline{5-6} \\
\colhead{Line} & \colhead{(eV)} & \colhead{log $gf$} & \colhead{(m\AA)} & \colhead{NLTE\tablenotemark{b}} &
\colhead{LTE\tablenotemark{a}}}
\startdata
Ne\,{\sc i} $\lambda 5852.488$ & 16.850 & $-$0.490 & 155 &       & 9.36  \\
Ne\,{\sc i} $\lambda 5881.895$ & 16.620 & $-$0.770 & 128 &       & 9.46  \\
Ne\,{\sc i} $\lambda 6029.997$ & 16.670 & $-$1.040 & 81  &       & 9.49  \\
Ne\,{\sc i} $\lambda 6074.338$ & 16.670 & $-$0.500 & 159 &       & 9.37  \\
%Ne\,{\sc i} $\lambda 6096.163$ & 16.670 & $-$0.310 &     &       &       \\
Ne\,{\sc i} $\lambda 6143.063$ & 16.620 & $-$0.100 & 285 &       & 9.52  \\
Ne\,{\sc i} $\lambda 6163.594$ & 16.710 & $-$0.620 & 137 &       & 9.40  \\
Ne\,{\sc i} $\lambda 6217.281$ & 16.620 & $-$0.960 & 84  &       & 9.42  \\
Ne\,{\sc i} $\lambda 6266.495$ & 16.710 & $-$0.370 & 222 &       & 9.55  \\
Ne\,{\sc i} $\lambda 6334.428$ & 16.620 & $-$0.320 & 241 &       & 9.57  \\
Ne\,{\sc i} $\lambda 6382.992$ & 16.670 & $-$0.240 & 254 &       & 9.56  \\
Ne\,{\sc i} $\lambda 6402.246$ & 16.620 & $+$0.330 & 409 &       & 9.59  \\
Ne\,{\sc i} $\lambda 6506.528$ & 16.670 & $-$0.030 & 287 &       & 9.49  \\
Ne\,{\sc i} $\lambda 6532.882$ & 16.710 & $-$0.720 & 123 &       & 9.47  \\
Ne\,{\sc i} $\lambda 6598.953$ & 16.850 & $-$0.360 & 119 &       & 9.11  \\
Ne\,{\sc i} $\lambda 6717.043$ & 16.850 & $-$0.360 & 138 &       & 9.22  \\
Ne\,{\sc i} $\lambda 7032.413$ & 16.620 & $-$0.260 & 229 &       & 9.47  \\
Mean... & \nodata & \nodata & \nodata &    8.67       & 9.40$\pm$0.13 \\
\enddata
\tablenotetext{a}{($T_{\rm eff}$, $\log g$, $\xi$)=(18300, 2.20, 16.0)}
\tablenotetext{b}{By applying correction on the LTE neon abundance}
\end{deluxetable}

\begin{deluxetable}{lccrcc}
\tabletypesize{\scriptsize}
\tablewidth{0pt}
\tablecolumns{7}
%\tablewidth{0pc}
\setcounter{table} {3}
\tablecaption{Photospheric line by line LTE neon abundances,
NLTE neon abundance, and
the line's measured equivalent width ($W_{\lambda}$) in m\AA\, for V1920,Cyg}
\tablehead{
\colhead{} & \colhead{$\chi$} & \colhead{} & \colhead{$W_{\lambda}$} &
\multicolumn{2}{c}{log $\epsilon(\rm Ne)$}\\
\cline{5-6} \\
\colhead{Line} & \colhead{(eV)} & \colhead{log $gf$} & \colhead{(m\AA)} & \colhead{NLTE\tablenotemark{b}} &
\colhead{LTE\tablenotemark{a}}}
\startdata
Ne\,{\sc i} $\lambda 5852.488$ & 16.850 & $-$0.490 & 185 &       & 9.18  \\
Ne\,{\sc i} $\lambda 5881.895$ & 16.620 & $-$0.770 & 204 &       & 9.47  \\
Ne\,{\sc i} $\lambda 6029.997$ & 16.670 & $-$1.040 & 114 &       & 9.38  \\
Ne\,{\sc i} $\lambda 6074.338$ & 16.670 & $-$0.500 & 230 &       & 9.34  \\
%Ne\,{\sc i} $\lambda 6096.163$ & 16.670 & $-$0.310 &     &       &       \\
Ne\,{\sc i} $\lambda 6143.063$ & 16.620 & $-$0.100 & 359 &       & 9.44  \\
Ne\,{\sc i} $\lambda 6163.594$ & 16.710 & $-$0.620 & 198 &       & 9.35  \\
Ne\,{\sc i} $\lambda 6217.281$ & 16.620 & $-$0.960 & 93  &       & 9.18  \\
Ne\,{\sc i} $\lambda 6266.495$ & 16.710 & $-$0.370 & 264 &       & 9.38  \\
Ne\,{\sc i} $\lambda 6334.428$ & 16.620 & $-$0.320 & 258 &       & 9.28  \\
Ne\,{\sc i} $\lambda 6382.992$ & 16.670 & $-$0.240 & 319 &       & 9.45  \\
Ne\,{\sc i} $\lambda 6402.246$ & 16.620 & $+$0.330 & 472 &       & 9.47  \\
Ne\,{\sc i} $\lambda 6506.528$ & 16.670 & $-$0.030 & 373 &       & 9.46  \\
Ne\,{\sc i} $\lambda 6532.882$ & 16.710 & $-$0.720 & 176 &       & 9.40  \\
Ne\,{\sc i} $\lambda 6598.953$ & 16.850 & $-$0.360 & 177 &       & 9.07  \\
Ne\,{\sc i} $\lambda 6717.043$ & 16.850 & $-$0.360 & 177 &       & 9.09  \\
Ne\,{\sc i} $\lambda 7032.413$ & 16.620 & $-$0.260 & 340 &       & 9.55  \\
Mean... & \nodata & \nodata & \nodata &    8.50       & 9.30$\pm$0.14 \\
\enddata
\tablenotetext{a}{($T_{\rm eff}$, $\log g$, $\xi$)=(16330, 1.80, 20.0)}
\tablenotetext{b}{By applying correction on the LTE neon abundance}
\end{deluxetable}

\begin{deluxetable}{lccrcc}
\tabletypesize{\scriptsize}
\tablewidth{0pt}
\tablecolumns{7}
%\tablewidth{0pc}
\setcounter{table} {4}
\tablecaption{Photospheric line by line LTE neon abundances,
NLTE neon abundance, and
the line's measured equivalent width ($W_{\lambda}$) in m\AA\, for HD\,124448}
\tablehead{
\colhead{} & \colhead{$\chi$} & \colhead{} & \colhead{$W_{\lambda}$} &
\multicolumn{2}{c}{log $\epsilon(\rm Ne)$}\\
\cline{5-6} \\
\colhead{Line} & \colhead{(eV)} & \colhead{log $gf$} & \colhead{(m\AA)} & \colhead{NLTE\tablenotemark{b}} &
\colhead{LTE\tablenotemark{a}}}
\startdata
Ne\,{\sc i} $\lambda 6334.428$ & 16.620 & $-$0.320 & 118 &       & 8.45  \\
Ne\,{\sc i} $\lambda 6717.043$ & 16.850 & $-$0.360 & 105 &       & 8.51  \\
Mean... & \nodata & \nodata & \nodata &    7.70       & 8.50$\pm$0.04 \\
\enddata
\tablenotetext{a}{($T_{\rm eff}$, $\log g$, $\xi$)=(15500, 1.90, 12.0)}
\tablenotetext{b}{By applying correction on the LTE neon abundance}
\end{deluxetable}

\begin{deluxetable}{lccrcc}
\tabletypesize{\scriptsize}
\tablewidth{0pt}
\tablecolumns{7}
%\tablewidth{0pc}
\setcounter{table} {5}
\tablecaption{Photospheric line by line LTE neon abundances,
NLTE neon abundance, and
the line's measured equivalent width ($W_{\lambda}$) in m\AA\, for PV\,Tel}
\tablehead{
\colhead{} & \colhead{$\chi$} & \colhead{} & \colhead{$W_{\lambda}$} &
\multicolumn{2}{c}{log $\epsilon(\rm Ne)$}\\
\cline{5-6} \\
\colhead{Line} & \colhead{(eV)} & \colhead{log $gf$} & \colhead{(m\AA)} & \colhead{NLTE\tablenotemark{b}} &
\colhead{LTE\tablenotemark{a}}}
\startdata
Ne\,{\sc i} $\lambda 5881.895$ & 16.620 & $-$0.770 & 233 &       & 8.52  \\
Ne\,{\sc i} $\lambda 6074.338$ & 16.670 & $-$0.500 & 324 &       & 8.53  \\
Ne\,{\sc i} $\lambda 6096.163$ & 16.670 & $-$0.310 & 354 &       & 8.42  \\
Ne\,{\sc i} $\lambda 6402.246$ & 16.620 & $+$0.330 & 671 &       & 8.70  \\
Ne\,{\sc i} $\lambda 7032.413$ & 16.620 & $-$0.260 & 422 &       & 8.58  \\
Mean... & \nodata & \nodata & \nodata &   7.60        & 8.53$\pm$0.08 \\
\enddata
\tablenotetext{a}{($T_{\rm eff}$, $\log g$, $\xi$)=(13750, 1.60, 25.0)}
\tablenotetext{b}{By applying correction on the LTE neon abundance}
\end{deluxetable}

\begin{deluxetable}{lccrcc}
\tabletypesize{\scriptsize}
\tablewidth{0pt}
\tablecolumns{6}
%\tablewidth{0pc}
\setcounter{table} {6}
\tablecaption{Photospheric line by line NLTE and LTE abundances, and
the line's measured equivalent width ($W_{\lambda}$) in m\AA\, for
LS\,IV\,+6$^{\circ}$\,002}
\tablehead{
\colhead{} & \colhead{$\chi$} & \colhead{} & \colhead{$W_{\lambda}$} &
\multicolumn{2}{c}{log $\epsilon(\rm X)$}\\
\cline{5-6} \\
\colhead{Line} & \colhead{(eV)} & \colhead{log $gf$} & \colhead{(m\AA)} &
\colhead{NLTE\tablenotemark{a}} &
\colhead{LTE\tablenotemark{b}}}
\startdata
H\,{\sc i} $\lambda 4340.462$ & 10.199 & $-$0.447 &  20 & 8.15 &  7.95 \\
H\,{\sc i} $\lambda 4861.323$ & 10.199 & $-$0.020 &  11 & 7.80 &  7.34 \\
        &         &         &         &               &               \\
Mean... & \nodata & \nodata & \nodata & 7.98$\pm$0.25   & 7.65$\pm$0.43 \\
        &         &         &         &               &               \\
C\,{\sc ii} $\lambda 4267.001$ & 18.046 & $+$0.563 &    &      &      \\
C\,{\sc ii} $\lambda 4267.183$ & 18.046 & $+$0.716 &    &      &      \\
C\,{\sc ii} $\lambda 4267.261$ & 18.046 & $-$0.584 & 331 & 8.70? &   8.90?  \\
        &         &         &         &               &               \\
C\,{\sc ii} $\lambda 4285.703$ & 24.602 & $-$0.430\tablenotemark{c} &  61 & 9.42  &   9.57 \\
        &         &         &         &               &               \\
C\,{\sc ii} $\lambda 4291.815$ & 24.603 & $-$0.500\tablenotemark{d} &     & &    \\
C\,{\sc ii} $\lambda 4291.858$ & 24.603 & $-$0.500\tablenotemark{d} & 107 & 9.67 &   10.18? \\
        &         &         &         &               &               \\
C\,{\sc ii} $\lambda 4313.106$ & 23.116 & $-$0.373 & 85 & 9.31 &   9.50 \\
        &         &         &         &               &               \\
C\,{\sc ii} $\lambda 4317.265$ & 23.119 & $-$0.005 & 147 & 9.53  &  9.72  \\
C\,{\sc ii} $\lambda 4318.606$ & 23.114 & $-$0.407 & 64  & 9.12 &   9.30  \\
C\,{\sc ii} $\lambda 4372.375$ & 24.656 & $+$0.057\tablenotemark{c} & 162 & 9.96? &   10.12? \\
C\,{\sc ii} $\lambda 4413.271$ & 24.603 & $-$0.610 & 67  & 9.70 &   9.85  \\
        &         &         &         &               &               \\
Mean... & \nodata & \nodata & \nodata & 9.46$\pm$0.22  & 9.59$\pm$0.21 \\
        &         &         &         &               &               \\
C\,{\sc iii} $\lambda 4067.940$ & 39.923 & $+$0.720 & 154 & 9.25 &     8.85 \\
        &         &         &         &               &              \\
C\,{\sc iii} $\lambda 4068.916$ & 39.924 & $+$0.838 &     &      &          \\
C\,{\sc iii} $\lambda 4068.916$ & 39.924 & $-$0.340 & 155 & 9.11 &     8.71 \\
        &         &         &         &               &               \\
C\,{\sc iii} $\lambda 4070.260$ & 39.925 & $+$0.953 &     &      &     \\
C\,{\sc iii} $\lambda 4070.306$ & 39.925 & $-$0.339 & 170 & 9.10 &     8.70 \\
        &         &         &         &               &               \\
C\,{\sc iii} $\lambda 4186.900$ & 40.010 & $+$0.918 & 181 & 9.29 &     8.88 \\
C\,{\sc iii} $\lambda 4647.418$ & 29.535 & $+$0.070 & 299 & 9.24 &     9.08 \\
C\,{\sc iii} $\lambda 4650.246$ & 29.535 & $-$0.151 & 262 & 9.30 &     9.10 \\
C\,{\sc iii} $\lambda 4651.473$ & 29.535 & $-$0.629 & 168 & 9.21 &     8.85 \\
C\,{\sc iii} $\lambda 4659.058$ & 38.218 & $-$0.654 & 59  & 9.60 &     9.18 \\
C\,{\sc iii} $\lambda 4663.642$ & 38.219 & $-$0.530 & 63  & 9.53 &     9.12 \\
C\,{\sc iii} $\lambda 4665.860$ & 38.226 & $+$0.044 & 101 & 9.40 &     9.01 \\
C\,{\sc iii} $\lambda 4673.953$ & 38.226 & $-$0.433 & 69  & 9.52 &     9.11 \\
        &         &         &         &               &               \\
Mean... & \nodata & \nodata & \nodata & 9.32$\pm$0.17 &  8.96$\pm$0.17 \\
        &         &         &         &               &               \\
N\,{\sc ii} $\lambda 3994.997$ & 18.497 & $+$0.208 & 139  & 7.88 &  8.34 \\
N\,{\sc ii} $\lambda 4035.081$ & 23.124 & $+$0.623\tablenotemark{e} &  97  & 8.01 &  8.27     \\
N\,{\sc ii} $\lambda 4041.310$ & 23.142 & $+$0.853\tablenotemark{e} & 129  & 8.10 &  8.37     \\
N\,{\sc ii} $\lambda 4043.532$ & 23.132 & $+$0.743\tablenotemark{e} &  92  & 7.84 &  8.10     \\
N\,{\sc ii} $\lambda 4044.779$ & 23.132 & $-$0.437\tablenotemark{e} &  66  & 8.74 &  8.99     \\
N\,{\sc ii} $\lambda 4056.907$ & 23.142 & $-$0.437\tablenotemark{e} &  49  & 8.53 &  8.78     \\
N\,{\sc ii} $\lambda 4073.053$ & 23.124 & $-$0.160\tablenotemark{d} &  62  & 8.42 &  8.67     \\
N\,{\sc ii} $\lambda 4082.270$ & 23.132 & $+$0.150\tablenotemark{d} &  70  & 8.21 &  8.46     \\
N\,{\sc ii} $\lambda 4173.561$ & 23.242 & $-$0.570\tablenotemark{d} &  42  & 8.60 &  8.85     \\
N\,{\sc ii} $\lambda 4179.674$ & 23.246 & $-$0.090\tablenotemark{d} &  55  & 8.30 &  8.55     \\
        &         &         &         &               &               \\
N\,{\sc ii} $\lambda 4236.927$ & 23.239 & $+$0.383\tablenotemark{e} &      & &           \\
N\,{\sc ii} $\lambda 4237.047$ & 23.242 & $+$0.553\tablenotemark{d} & 109  & 7.93 &  8.18     \\
        &         &         &         &               &               \\
N\,{\sc ii} $\lambda 4427.233$ & 23.422 & $-$0.010\tablenotemark{d} &  58  & 8.33 &  8.57     \\
N\,{\sc ii} $\lambda 4427.963$ & 23.422 & $-$0.170\tablenotemark{e} &  60  & 8.50 &  8.74     \\
N\,{\sc ii} $\lambda 4431.814$ & 23.415 & $-$0.170\tablenotemark{e} &  47  & 8.33 &  8.57     \\
N\,{\sc ii} $\lambda 4432.736$ & 23.415 & $+$0.580\tablenotemark{e} &  92  & 8.09 &  8.34     \\
N\,{\sc ii} $\lambda 4433.475$ & 23.425 & $-$0.040\tablenotemark{e} &  58  & 8.35 &  8.58     \\
N\,{\sc ii} $\lambda 4442.015$ & 23.422 & $+$0.310\tablenotemark{e} &  50  & 7.89 &  8.13     \\
N\,{\sc ii} $\lambda 4447.030$ & 20.409 & $+$0.228 & 110  & 8.03 &  8.40 \\
N\,{\sc ii} $\lambda 4530.410$ & 23.475 & $+$0.670\tablenotemark{e} & 103  & 8.16 &  8.41     \\
N\,{\sc ii} $\lambda 4601.478$ & 18.466 & $-$0.428 & 112  & 8.30 &  8.74 \\
N\,{\sc ii} $\lambda 4607.153$ & 18.462 & $-$0.507 & 118  & 8.45 &  8.89 \\
N\,{\sc ii} $\lambda 4613.868$ & 18.466 & $-$0.665 &  93  & 8.34 &  8.76 \\
N\,{\sc ii} $\lambda 4621.393$ & 18.466 & $-$0.514 & 122  & 8.50 &  8.94 \\
N\,{\sc ii} $\lambda 4630.539$ & 18.483 & $+$0.094 & 173  & 8.45 &  8.97 \\
N\,{\sc ii} $\lambda 4678.135$ & 23.572 & $+$0.434\tablenotemark{e} &  66  & 8.04 &  8.27     \\
N\,{\sc ii} $\lambda 4694.642$ & 23.572 & $+$0.100\tablenotemark{d} &  65  & 8.37 &  8.61     \\
N\,{\sc ii} $\lambda 4774.244$ & 20.646 & $-$1.257 &  27  & 8.55 &  8.87 \\
N\,{\sc ii} $\lambda 4779.722$ & 20.646 & $-$0.587 &  58  & 8.35 &  8.68 \\
N\,{\sc ii} $\lambda 4781.190$ & 20.654 & $-$1.308 &  19  & 8.43 &  8.74 \\
N\,{\sc ii} $\lambda 4793.648$ & 20.654 & $-$1.095 &  33  & 8.51 &  8.83 \\
N\,{\sc ii} $\lambda 4803.287$ & 20.666 & $-$0.113 &  88  & 8.23 &  8.57 \\
N\,{\sc ii} $\lambda 4810.299$ & 20.666 & $-$1.084 &  26  & 8.37 &  8.68 \\
        &         &         &         &               &               \\
Mean... & \nodata & \nodata & \nodata & 8.29$\pm$0.23 & 8.59$\pm$0.25 \\
        &         &         &         &               &               \\
N\,{\sc iii} $\lambda 4097.360$ & 27.438 & $-$0.057 & 184  & 8.23 &  8.39 \\
N\,{\sc iii} $\lambda 4103.390$ & 27.438 & $-$0.359 & 169 & 8.44 &  8.56 \\
N\,{\sc iii} $\lambda 4215.770$ & 36.856 & $-$0.705 &  18 & 8.41 &  8.54 \\
N\,{\sc iii} $\lambda 4379.201$ & 39.711 & $+$1.010\tablenotemark{c} & 133 & 8.31 &  8.90     \\
N\,{\sc iii} $\lambda 4514.850$ & 35.671 & $+$0.221 & 112 & 9.15? &  9.10?  \\
N\,{\sc iii} $\lambda 4518.140$ & 35.649 & $-$0.461 &  75 & 9.40? &  9.33?  \\
N\,{\sc iii} $\lambda 4523.560$ & 35.657 & $-$0.353 &  51 & 8.91? &  8.86 \\
N\,{\sc iii} $\lambda 4527.860$ & 38.958 & $-$0.471\tablenotemark{c} &  23 & 8.41 &  8.88    \\
N\,{\sc iii} $\lambda 4535.050$ & 38.958 & $-$0.170\tablenotemark{c} &  26 & 8.19 &  8.66    \\
N\,{\sc iii} $\lambda 4539.700$ & 38.645 & $-$0.452 &  33 & 8.64 &  9.07?   \\
N\,{\sc iii} $\lambda 4544.840$ & 39.396 & $-$0.151 &  33 & 8.34 &  8.77    \\
N\,{\sc iii} $\lambda 4546.330$ & 38.958 & $+$0.004\tablenotemark{c} &  21 & 7.90 &  8.35    \\
N\,{\sc iii} $\lambda 4634.130$ & 30.459 & $-$0.086 & 115 & 8.63 &  8.47 \\
N\,{\sc iii} $\lambda 4640.640$ & 30.463 & $+$0.168 & 131 & 8.50 &  8.37 \\
N\,{\sc iii} $\lambda 4641.850$ & 30.463 & $-$0.788 & 139 & 9.53? &  9.41?  \\
        &         &         &         &               &               \\
Mean... & \nodata & \nodata & \nodata & 8.36$\pm$0.21 & 8.61$\pm$0.21 \\
        &         &         &         &               &               \\
O\,{\sc ii} $\lambda 4069.623$ & 25.631 & $+$0.150 &      &      &         \\
O\,{\sc ii} $\lambda 4069.882$ & 25.638 & $+$0.344 & 140  & 8.25 &  8.18   \\
        &         &         &         &               &               \\
O\,{\sc ii} $\lambda 4072.153$ & 25.650 & $+$0.552 &  88  & 8.04 &  7.96    \\
O\,{\sc ii} $\lambda 4085.112$ & 25.650 & $-$0.189 &  49  & 8.68 &  8.22    \\
O\,{\sc ii} $\lambda 4092.929$ & 25.665 & $-$0.308 &  38  & 8.40 &  8.18    \\
O\,{\sc ii} $\lambda 4366.895$ & 22.999 & $-$0.348 & 115  & 8.39 &  8.72   \\
O\,{\sc ii} $\lambda 4414.899$ & 23.442 & $+$0.172 & 118  & 7.93 &  8.30   \\
O\,{\sc ii} $\lambda 4416.975$ & 23.419 & $-$0.077 & 134  & 8.31 &  8.73   \\
O\,{\sc ii} $\lambda 4452.378$ & 23.442 & $-$0.788 &  55  & 8.25 &  8.53   \\
O\,{\sc ii} $\lambda 4590.974$ & 25.661 & $+$0.350 &  99  & 8.43 &  8.36   \\
        &         &         &         &               &              \\
O\,{\sc ii} $\lambda 4595.957$ & 25.661 & $-$1.033 &      &      &     \\
O\,{\sc ii} $\lambda 4596.177$ & 25.661 & $+$0.200 &  87  & 8.38 &  8.30   \\
        &         &         &         &               &              \\
O\,{\sc ii} $\lambda 4638.856$ & 22.966 & $-$0.332 & 110  & 8.34 &  8.67   \\
O\,{\sc ii} $\lambda 4649.135$ & 22.999 & $+$0.308 & 151  & 8.08 &  8.52   \\
O\,{\sc ii} $\lambda 4661.632$ & 22.979 & $-$0.278 &  90  & 8.09 &  8.39   \\
O\,{\sc ii} $\lambda 4676.235$ & 22.999 & $-$0.394 &  67  & 7.95 &  8.23   \\
O\,{\sc ii} $\lambda 4696.353$ & 28.510 & $-$1.380 &  16  & 8.07 &  8.36   \\
        &         &         &         &               &               \\
Mean... & \nodata & \nodata & \nodata & 8.24$\pm$0.20  & 8.38$\pm$0.22 \\
        &         &         &         &               &               \\
%Ne\,{\sc ii} $\lambda 4217.066$ & 34.609 & $-$1.230\tablenotemark{d} &     &    & \\
Ne\,{\sc ii} $\lambda 4217.169$ & 34.609 & $+$0.090\tablenotemark{d} & 45  & 8.59  & 8.57 \\
%        &         &         &         &               &              \\
Ne\,{\sc ii} $\lambda 4219.745$ & 34.609 & $+$0.750\tablenotemark{d} & 99  & 8.60 &  8.64 \\
        &         &         &         &               &              \\
Ne\,{\sc ii} $\lambda 4220.894$ & 34.619 & $-$0.060\tablenotemark{d} &     & &  \\
Ne\,{\sc ii} $\lambda 4221.087$ & 34.619 & $-$0.740\tablenotemark{d} & 37  & 8.52 &  8.49 \\
        &         &         &         &               &              \\
Ne\,{\sc ii} $\lambda 4224.472$ & 34.632 & $-$0.860\tablenotemark{d} &     & &  \\
Ne\,{\sc ii} $\lambda 4224.642$ & 34.632 & $-$0.750\tablenotemark{d} & 23  & 8.77 &  8.75 \\
        &         &         &         &               &              \\
Ne\,{\sc ii} $\lambda 4231.532$ & 34.619 & $-$0.080\tablenotemark{d} &     & &  \\
Ne\,{\sc ii} $\lambda 4231.636$ & 34.619 & $+$0.260\tablenotemark{d} & 60  & 8.46 &  8.44 \\
        &         &         &         &               &              \\
Ne\,{\sc ii} $\lambda 4239.911$ & 34.632 & $-$0.490\tablenotemark{d} &     & &  \\
Ne\,{\sc ii} $\lambda 4240.105$ & 34.632 & $-$0.020\tablenotemark{d} & 50  & 8.62 &  8.59 \\
        &         &         &         &               &              \\
Ne\,{\sc ii} $\lambda 4250.645$ & 34.632 & $+$0.150\tablenotemark{d} & 46  & 8.58 &  8.56 \\
Ne\,{\sc ii} $\lambda 4397.991$ & 34.814 & $+$0.160\tablenotemark{d} & 61  & 8.91 &  8.82 \\
Ne\,{\sc ii} $\lambda 4430.946$ & 34.749 & $+$0.310\tablenotemark{d} & 72  & 8.78 &  8.81 \\
        &         &         &         &               &               \\
Mean... & \nodata & \nodata & \nodata & 8.65$\pm$0.14  & 8.63$\pm$0.14 \\
\enddata
\tablenotetext{a}{($T_{\rm eff}$, $\log g$, $\xi$)=(30000, 4.10, 9.0)}
\tablenotetext{b}{($T_{\rm eff}$, $\log g$, $\xi$)=(32000, 4.20, 9.0)}
\tablenotetext{c}{\citet{jeff98}}
\tablenotetext{d}{Kurucz $gf$-value}
\tablenotetext{e}{\citet{wiese66}}
\end{deluxetable}

\begin{deluxetable}{lccccccc}
\tabletypesize{\scriptsize}
\tablewidth{0pt}
\tablecolumns{8}
%\tablewidth{0pc}
\setcounter{table} {7}
\tablecaption{He, C, N, O, Ne, and Fe abundances of the sample EHe stars}
\tablehead{
\colhead{} & \colhead{} & \multicolumn{6}{c}{log $\epsilon(\rm X)$}\\
\cline{3-7} \\
\colhead{Star} & \colhead{($T_{\rm eff}$, $\log g$)} & \colhead{He} & \colhead{C} & \colhead{N} & \colhead{O} &
\colhead{Ne} & \colhead{Fe\tablenotemark{a}}}
\startdata
LS\,IV\,+6$^{\circ}$\,002 & (30000, 4.10) & 11.54 & 9.4 & 8.3 & 8.2 & 8.65 & 7.1 \\
BD$-9^\circ$4395 & (24300, 2.65) & 11.54 & 9.1 & 7.8 & 8.0 & 8.18 & 6.6 \\
LSE\,78 & (18300, 2.20) & 11.54 & 9.4 & 8.3 & 9.4 & 8.67 & 6.8 \\
BD+10$^\circ$2179 & (16375, 2.45) & 11.54 & 9.3 & 8.1 & 7.9 & 7.87 & 6.2 \\
V1920\,Cyg & (16330, 1.80) & 11.50 & 9.6 & 8.6 & 9.9 & 8.50 & 6.8 \\
HD\,124448 & (15500, 1.90) & 11.54 & 9.1 & 8.7 & 8.3 & 7.70 & 7.2 \\
PV\,Tel & (13750, 1.60) & 11.54 & 9.2 & 8.7 & 8.8 & 7.60 & 7.0 \\
\enddata
\tablenotetext{a}{$\rm LTE$ abundance}
\end{deluxetable}

\begin{figure}
\epsscale{1.00}
\plotone{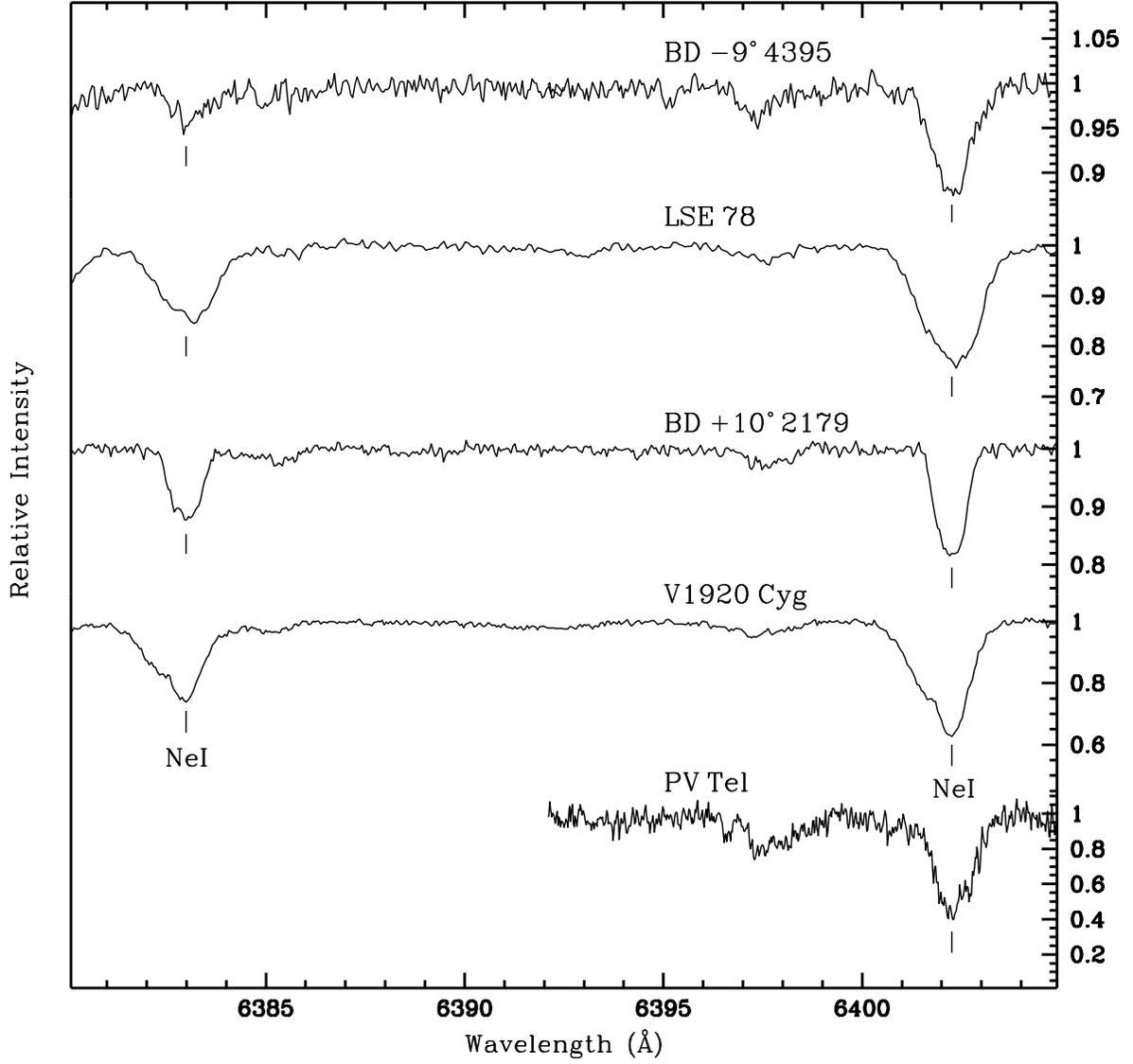}
\caption{The spectral region from 6380 -- 6405 \AA\ is shown for five EHes
with the
hottest star at the top and the coolest star at the bottom. The Ne\,{\sc i}
lines at 6382.99 \AA\ and at 6402.25 \AA\ are marked.
 \label{fig1}}
\end{figure}

\begin{figure}
\epsscale{1.00}
\plotone{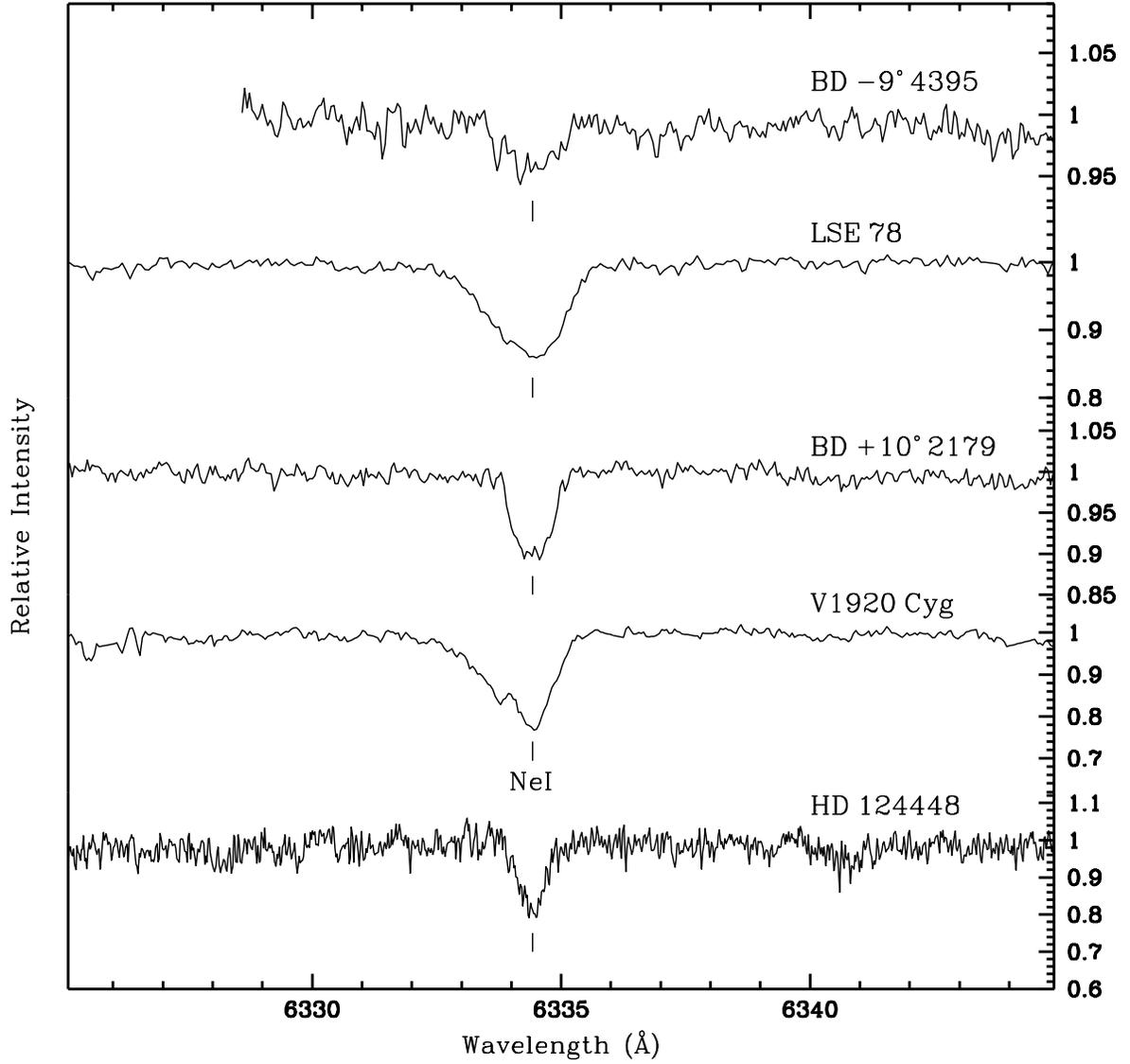}
\caption{The spectral region from 6325 -- 6345 \AA\ is shown for five EHes
with the
hottest star at the top and the coolest star at the bottom. The Ne\,{\sc i}
line at 6334.43 \AA\ is marked. \label{fig2}}
\end{figure}

\begin{figure}
\epsscale{1.00}
\plotone{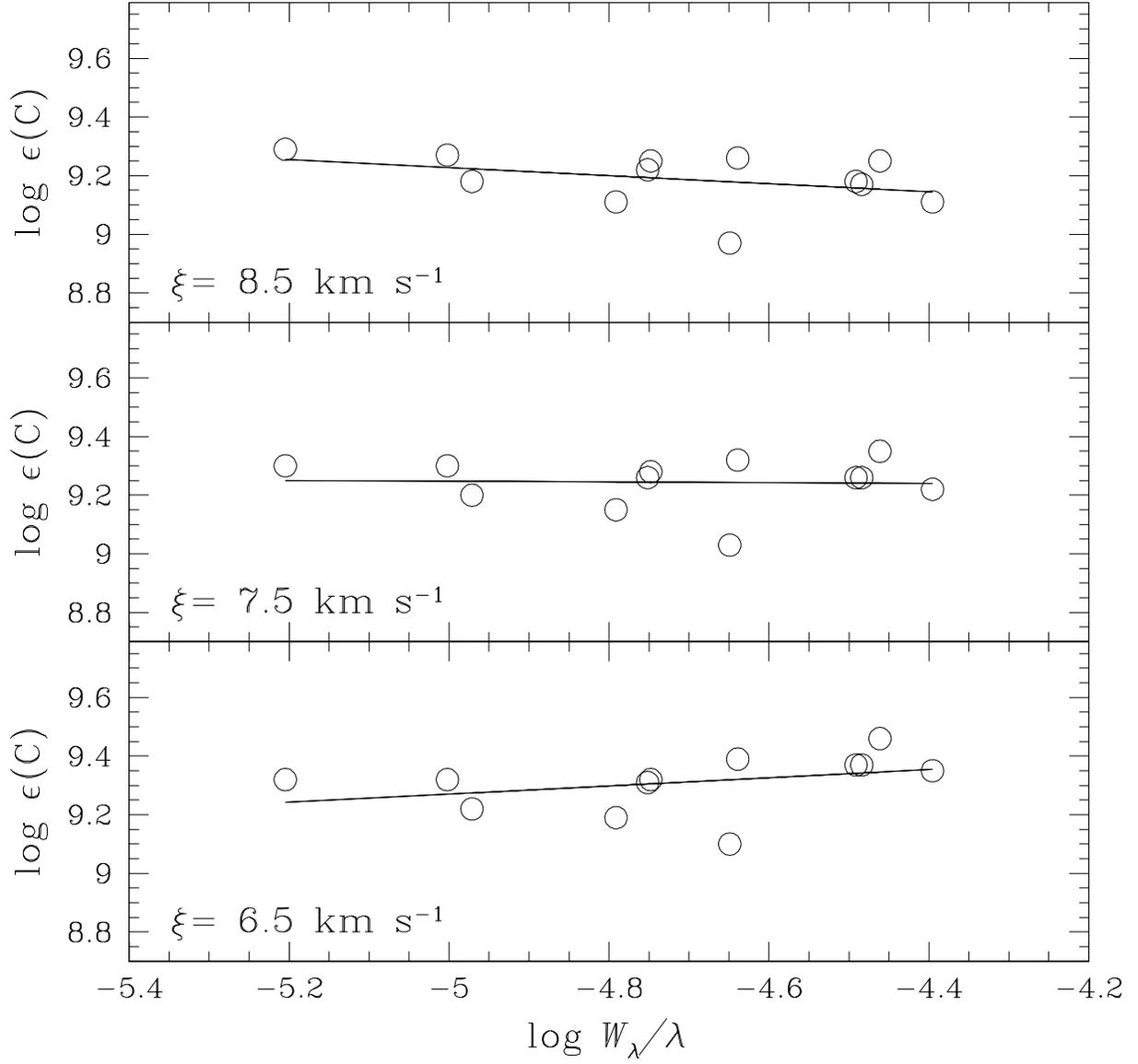}
\caption{NLTE abundances from C\,{\sc ii} lines for BD\,+10$^{\circ}$\,2179
versus their reduced equivalent widths ($\log {\it W_{\lambda}/\lambda}$).
A microturbulent velocity of $\xi$ = 7.5 km s$^{-1}$ is obtained from
this figure. \label{fig3}}
\end{figure}

\begin{figure}
\epsscale{1.00}
\plotone{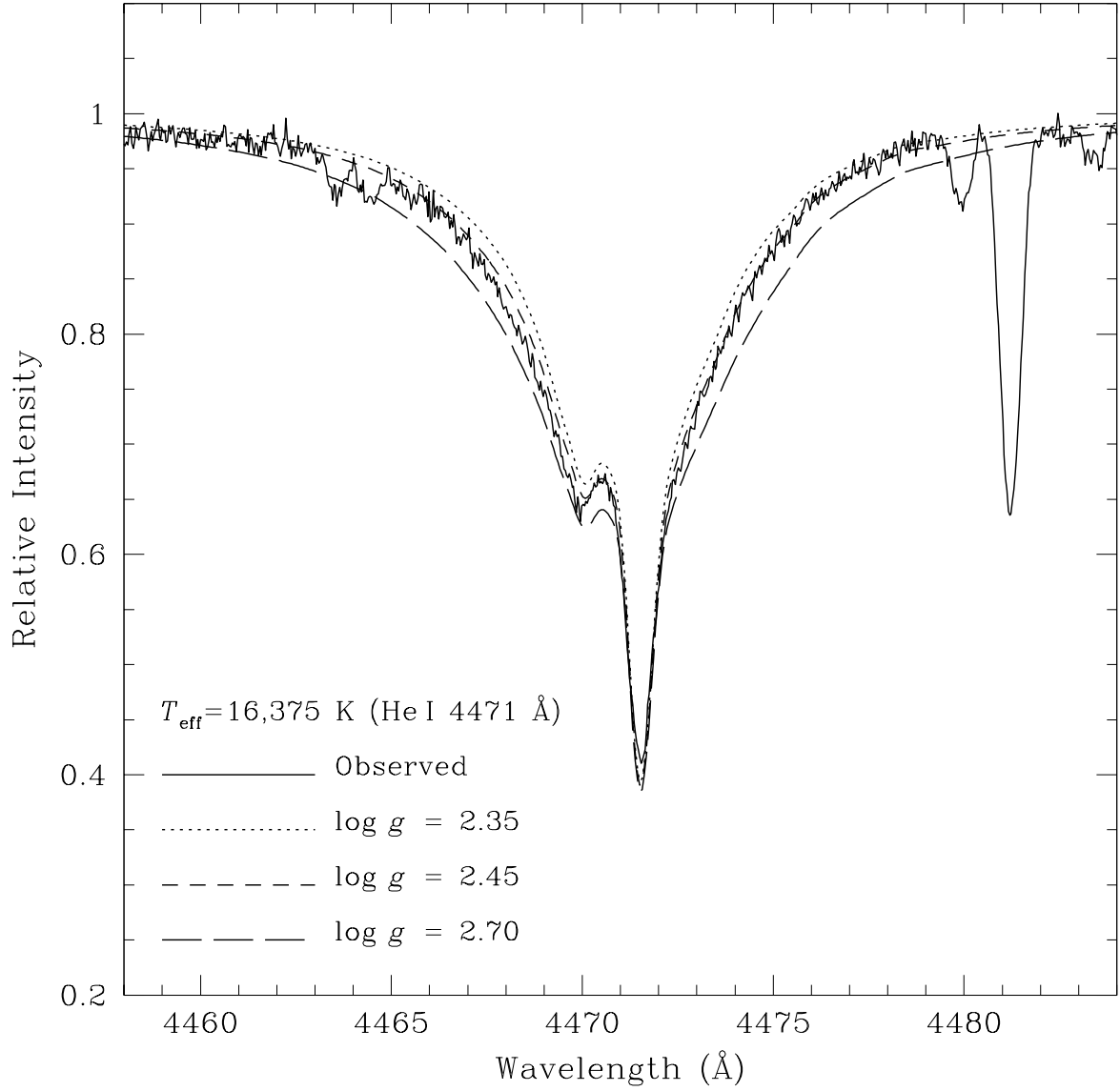}
\caption{BD\,+10$^{\circ}$\,2179's observed and synthesized NLTE He\,{\sc i}
line
profile at 4471 \AA. The NLTE He\,{\sc i} line profiles are synthesized using
the
NLTE model $T_{\rm eff}$=16,375 K, for three different $\log g$ values $-$ see
key
on the figure. \label{fig4}}
\end{figure}

\begin{figure}
\epsscale{1.00}
\plotone{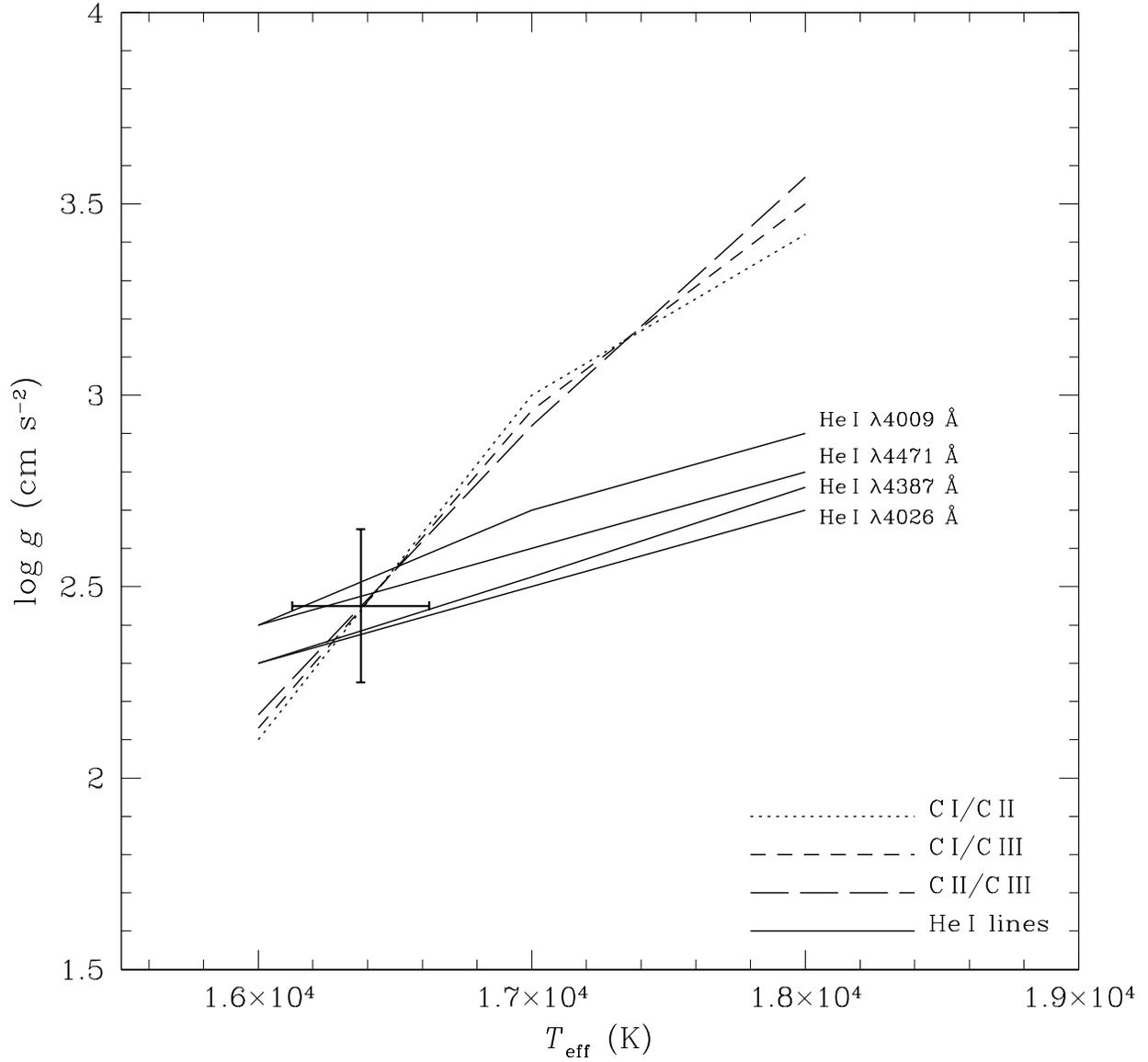}
\caption{The $T_{\rm eff}$ vs $\log g$ plane for BD\,+10$^{\circ}$\,2179.
Loci satisfying ionization equilibria are plotted $-$ see keys on the figure.
The loci satisfying  optical He\,{\sc i} line profiles ($\lambda$ 4471, 4387,
4026, and 4009 \AA)
are shown by the solid lines. The cross shows the adopted
NLTE model atmosphere parameters. \label{fig5}}
\end{figure}

%\begin{figure}
%\epsscale{1.00}
%\plotone{nltfig4.eps}
%\caption{NLTE abundances from O\,{\sc ii} lines for BD\,--9$^{\circ}$\,4395
%versus their reduced equivalent widths ($\log {\it W_{\lambda}/\lambda}$).
%A microturbulent velocity of $\xi$ = 17.5 km s$^{-1}$ is obtained from
%this figure. \label{fig6}}
%\end{figure}

\begin{figure}
\epsscale{1.00}
\plotone{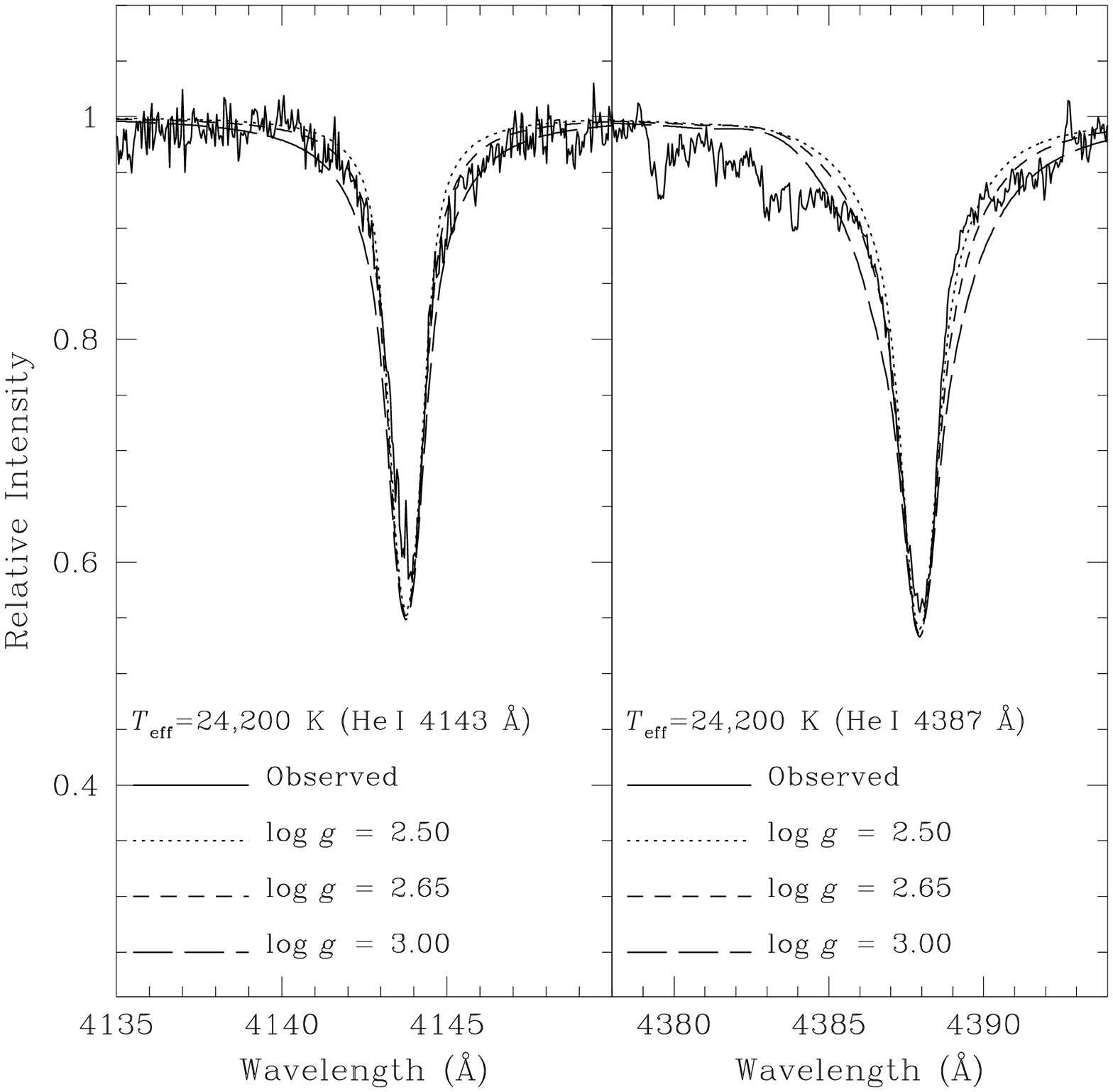}
\caption{BD\,--9$^{\circ}$\,4395's observed and synthesized NLTE He\,{\sc i} line
profiles at 4143 \AA\ and at 4387 \AA. The NLTE He\,{\sc i} line profiles are
synthesized using the
NLTE model $T_{\rm eff}$=24,200 K, for three different $\log g$ values $-$ see key
on the figure. \label{fig6}}
\end{figure}

\begin{figure}
\epsscale{1.00}
\plotone{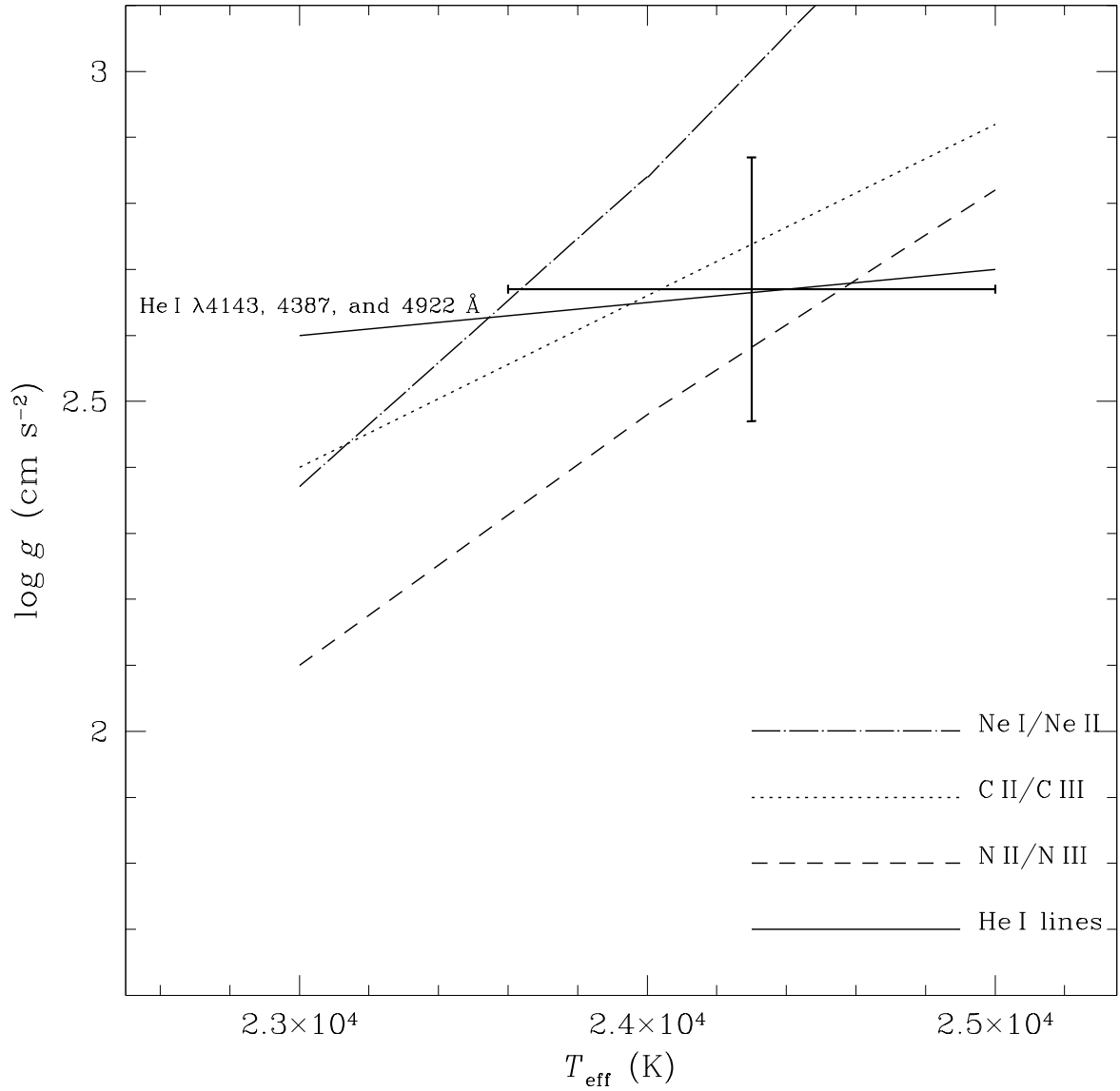}
\caption{The $T_{\rm eff}$ vs $\log g$ plane for BD\,--9$^{\circ}$\,4395.
Loci satisfying ionization equilibria are plotted $-$ see keys on the figure.
The locus satisfying  optical He\,{\sc i} line profiles ($\lambda$ 4143, 4387,
and 4922 \AA) is shown by the solid line. The cross shows the adopted
NLTE model atmosphere parameters. \label{fig7}}
\end{figure}

\begin{figure}
\epsscale{1.00}
\plotone{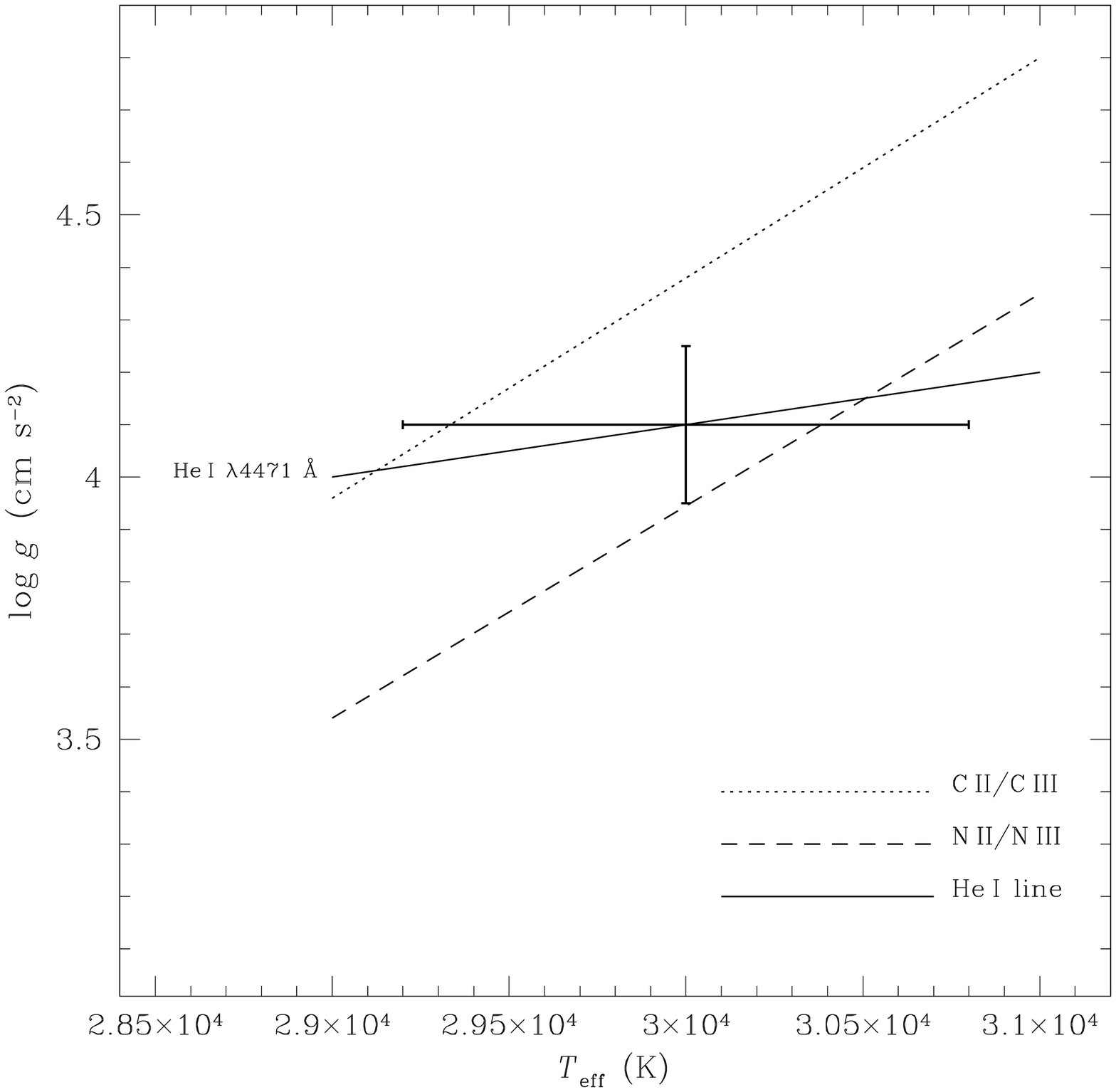}
\caption{The $T_{\rm eff}$ vs $\log g$ plane for LS\,IV\,+6$^{\circ}$\,002.
Loci satisfying ionization equilibria are plotted $-$ see keys on the figure.
The locus satisfying  optical He\,{\sc i} 4471 \AA\ line profile 
is shown by the solid line. The cross shows the adopted
NLTE model atmosphere parameters. \label{fig8}}
\end{figure}

\begin{figure}
\epsscale{1.00}
\plotone{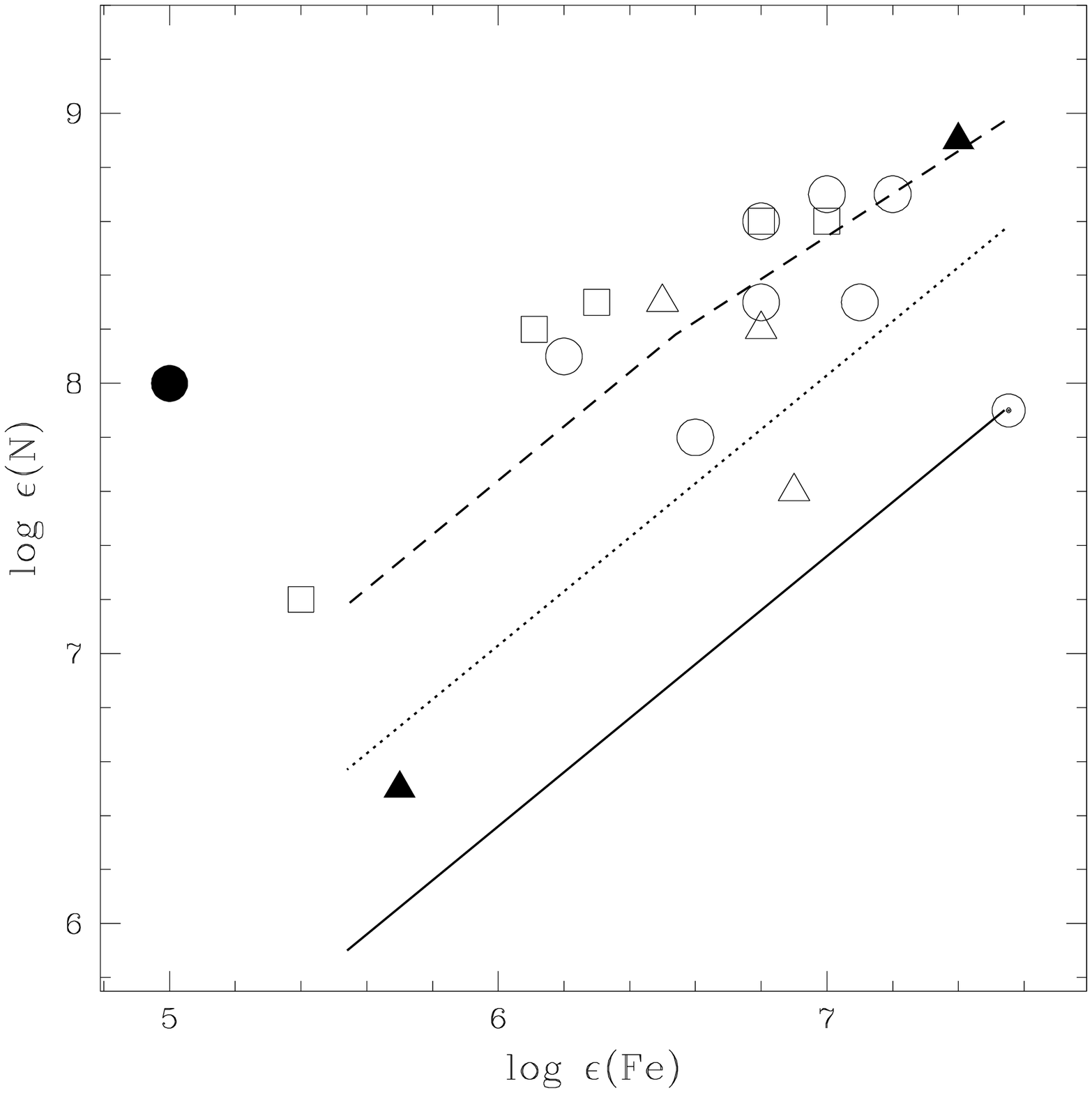}
\caption{N vs. Fe. Our sample of seven EHes is represented by open circles.
Five cool EHes are represented by open squares \citep{pan2006,panred2006,pan2001}.
The results taken from the literature for  EHes with
C/He of about 1\% \citep{drill98,jef98} are represented by open triangles. The two
EHes of much lower C/He -- V652\,Her and HD\,144941 -- are shown by
filled triangles \citep{jeff97,har1997,jeff99}.
DY\,Cen \citep{jefheb93} is represented by a filled circle. Circled dot
represents the Sun. N = Fe is denoted by the solid line. The dotted line
represents conversion of the initial sum of C and N to N. The dashed line
represents the locus of the sum of initial C, N, and O converted to N. \label{fig9}}
\end{figure}

\begin{figure}
\epsscale{1.00}
\plotone{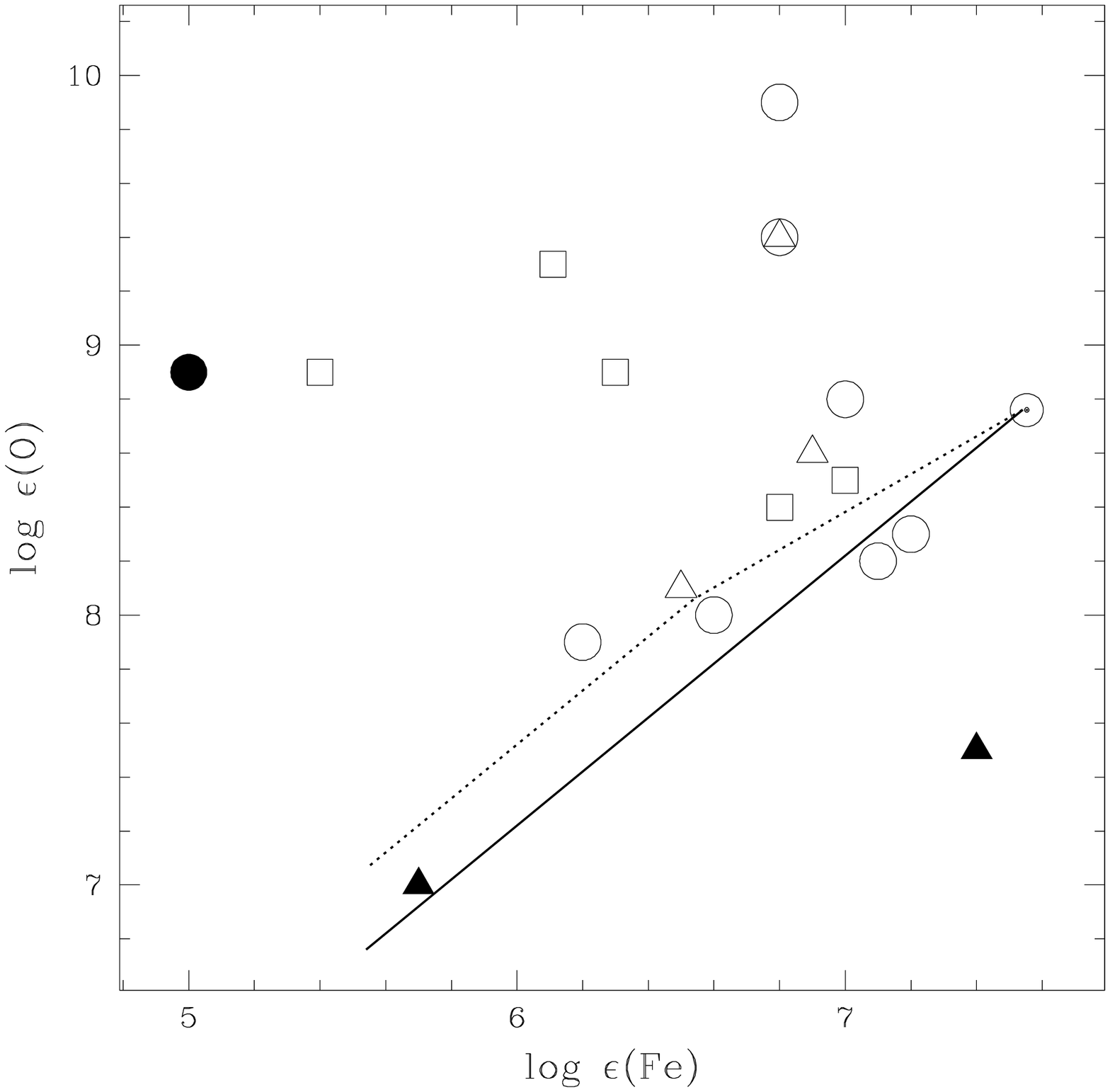}
\caption{O vs. Fe. Our sample of seven EHes is represented by open circles.
Five cool EHes are represented by open squares \citep{pan2006,panred2006,pan2001}.
The results taken from the literature for  EHes with
C/He of about 1\% \citep{drill98,jef98} are represented by open triangles. The two
EHes of much lower C/He -- V652\,Her and HD\,144941 -- are shown by
filled triangles \citep{jeff97,har1997,jeff99}.
DY\,Cen \citep{jefheb93} is represented by a filled circle. Circled dot
represents the Sun. O = Fe is denoted by the solid line. The dotted line
is from the relation [$\alpha$/Fe] vs. [Fe/H] for normal
disk and halo stars \citep{rydelamb2004}. \label{fig10}}
\end{figure}

\begin{figure}
\epsscale{1.00}
\plotone{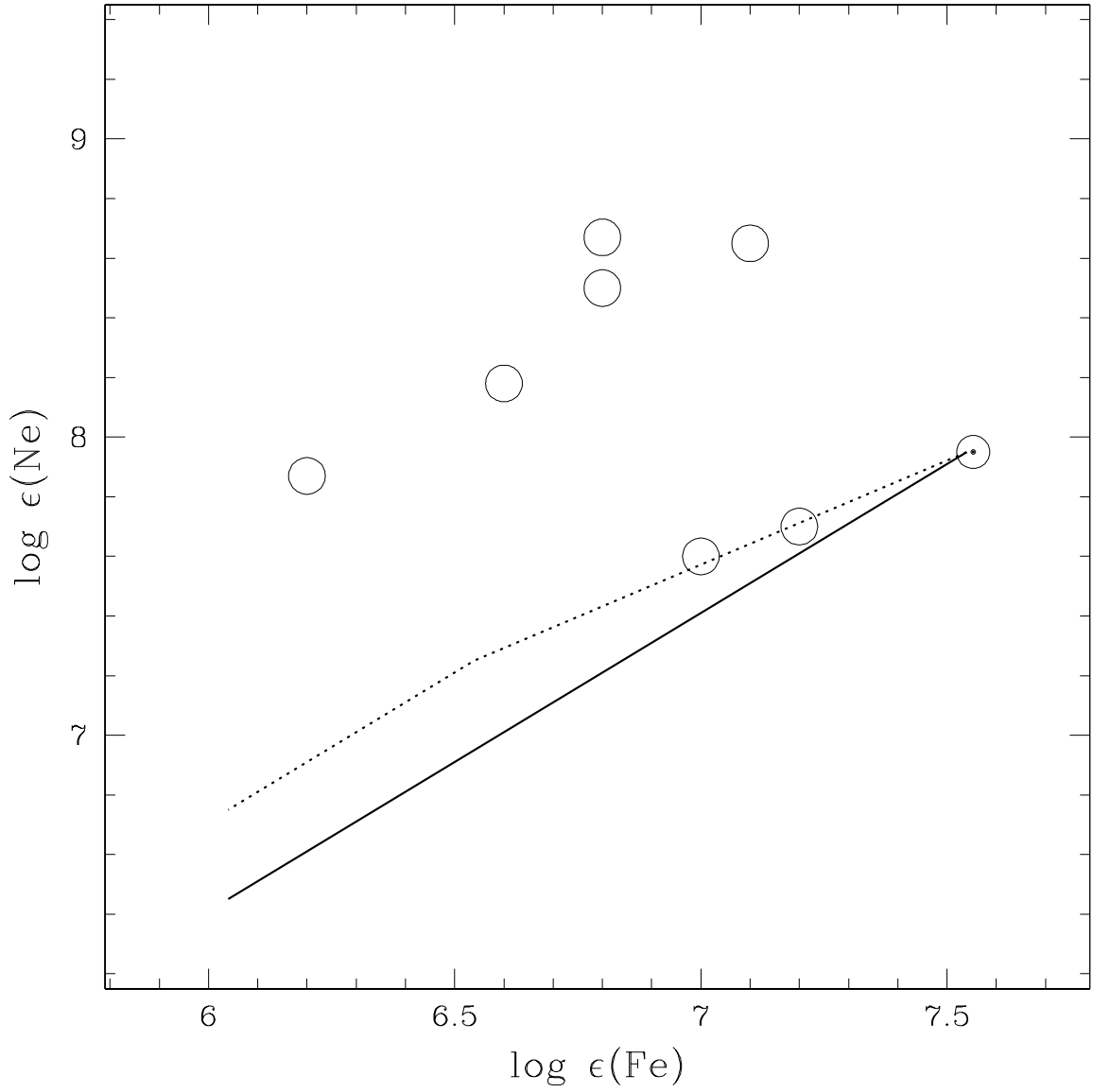}
\caption{Ne vs. Fe. Our sample of seven EHe stars is represented by circles.
The symbol $\odot$ is the Sun. Ne = Fe is denoted by the solid line.
The dotted line is from the relation [$\alpha$/Fe] vs. [Fe/H] for normal
disk and halo stars \citep{rydelamb2004}. \label{fig11}}
\end{figure}

\begin{figure}
\epsscale{1.00}
\plotone{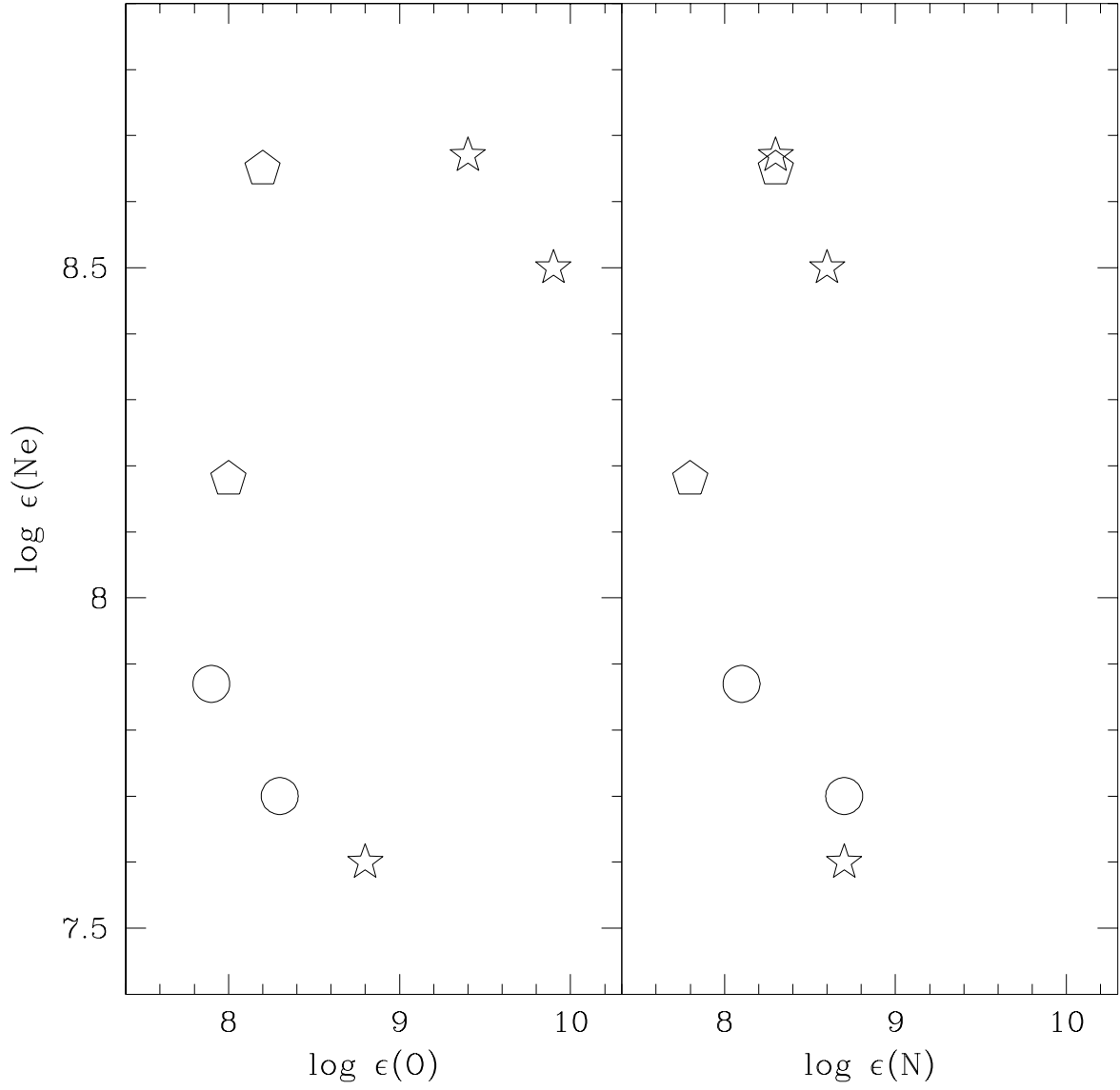}
\caption{$Left$: Ne vs. O. Our sample of seven EHe stars. $s$-process enriched EHes
are represented by stars. EHes with no $s$-process enrichment are represented
by circles. EHes with no information on $s$-process enrichment are represented
by pentagons. $Right$: Ne vs. N. The symbols have the same meaning as in the 
left panel. \label{fig12}}
\end{figure}


\clearpage

\begin{thebibliography}{}
\bibitem[Abia et al.(1999)]{abia99} Abia, C., Pavlenko, Y., \& de Laverny, P. 1999, A\&A, 351, 273
\bibitem[Asplund et al.(2000)]{asp00} Asplund, M., Gustafsson, B., Lambert, D. L.,
  \& Rao, N. K. 2000, \aap, 353, 287
\bibitem[Auer \& Mihalas(1973)]{aur73} Auer, L. H., \& Mihalas, D. 1973, \apj, 184, 151
%\bibitem[Barnard, Cooper \& Smith(1974)]{barn74}  Barnard, A. J., Cooper, J.,
%\& Smith, E. W. 1974, JQSRT, 14, 1025
\bibitem[Bl\"{o}cker(2001)]{bloc2001} Bl\"{o}cker, T. 2001, \apss, 275, 1 
\bibitem[Clayton et al.(2005)]{clay2005} Clayton, G. C., Herwig, F., Geballe, T. R., Asplund, M.,
Tenenbaum, E. D., Engelbracht, C. W., \& Gordon, K. D. 2005, \apjl, 623, L141
\bibitem[Clayton et al.(2007)]{clay2007} Clayton, G. C., Geballe, T. R., Herwig, F., Fryer, C.,
\& Asplund, M. 2007 \apj, 662, 1220
%\bibitem[Cunha \& Lambert(1992)]{cl92} Cunha, K., \& Lambert, D. L. 1992,
%\apj, 399, 586
\bibitem[Cunha, Hubeny \& Lanz(2006)]{chl06} Cunha, K., Hubeny, I., \& Lanz, T.
2006, \apjl, 647, L143 
\bibitem[Drilling, Jeffery, \& Heber(1998)]{drill98} Drilling, J. S., Jeffery C. S.,
\& Heber, U. 1998, \aap, 329, 1019
%\bibitem[Dworetsky \& Budaj(2000)]{dwb00} Dworetsky, M. M., \& Budaj, J. 
%2000, \mnras, 318, 1264
\bibitem[Cameron \& Fowler(1971)]{cam1971} Cameron, A.G.W., \& Fowler, W.A. 1971, ApJ, 164, 111
\bibitem[Garc\'{\i}a-Hern\'{a}ndez et al.(2009)]{garc2009} Garc\'{\i}a-Hern\'{a}ndez, D. A.,
Hinkle, K. H., Lambert, D. L., \& Eriksson, K. 2009, \apj, 696, 1733
\bibitem[Garc\'{\i}a-Hern\'{a}ndez et al.(2010)]{garc2010} Garc\'{\i}a-Hern\'{a}ndez, D. A.,
Lambert, D. L., Rao, N. K., Hinkle, K. H., \& Eriksson, K. 2010, \apj, 714, 144
%\bibitem[Goswami \& Prantzos(2000)]{GP2000} Goswami, A., \& Prantzos, N.
%2000, \aap, 359, 191
\bibitem[Harrison \& Jeffery(1997)]{har1997} Harrison, P.M., \& Jeffery, C.S. 1997, A\&A, 323, 177
%\bibitem[Heber(1983)]{heb83} Heber, U. 1983, \aap, 118, 39
\bibitem[Herwig(2000)]{herwig2000} Herwig, F. 2000, \aap, 360, 952
\bibitem[Herwig(2001)]{herwig2001} Herwig, F. 2001, \apss, 275, 15
\bibitem[Herwig(2006)]{herwig2006} Herwig, F. 2006, Proceedings of the
International Symposium on Nuclear Astrophysics - Nuclei in the Cosmos - IX.
25-30 June 2006, CERN, 206
\bibitem[Hubeny(1988)]{hub88} Hubeny, I. 1988, Comput. Phys. Commun., 52, 103
\bibitem[Hubeny, Hummer \& Lanz(1994)]{hhl1994} Hubeny, I., Hummer, D. G., \& Lanz, T. 1994, \aap, 282, 151
\bibitem[Hubeny \& Lanz(1995)]{hl95} Hubeny, I., \& Lanz, T. 1995, \apj, 439, 875
\bibitem[Hubeny, Lanz \& Jeffery(1995)]{hub95} Hubeny, I., Lanz, T.,
\& Jeffery, C. S. 1994, SYNSPEC: a User's Guide, in Newsl. on Analysis of
Astronomical Spectra 20 (St. Andrews Univ.)
\bibitem[Iben(1967)]{iben67} Iben, I., Jr. 1967, ApJ, 147, 624
\bibitem[Iben et al.(1983)]{iben83} Iben, I., Jr., Kaler, J. B., Truran, J. W., \& Renzini, A. 1983, 
\apj, 264, 605
\bibitem[Iben \& Tutukov(1984)]{ibandtutu84} Iben, I., Jr., \& Tutukov, A. V. 1984, \apjs, 54, 335 
\bibitem[Iben, Tutukov \& Yungelson(1996)]{iben96} Iben, I., Jr., Tutukov, A. V., \& Yungelson, L. R. 1996,
\apj, 456, 750
\bibitem[Iben, Tutukov \& Yungelson(1997)]{iben97} Iben, I. Jr., Tutukov, A. V., \& Yungelson, L. R. 1997,
\apj, 475, 291
\bibitem[Jeffery(1996)]{jeff96} Jeffery C. S. 1996, ASP Conf. Ser. 96, 152
\bibitem[Jeffery(1998)]{jeff98} Jeffery, C. S. 1998, \mnras, 294, 391
\bibitem[Jeffery \& Heber(1992)]{jefheb92} Jeffery, C. S., \& Heber, U. 1992,
\aap, 260, 133
\bibitem[Jeffery \& Heber(1993)]{jefheb93} Jeffery, C. S., \& Heber, U. 1993,
\aap, 270, 167
\bibitem[Jeffery \& Harrison(1997)]{jeff97} Jeffery, C. S., \& Harrison, P. M. 1997,
\aap, 323, 393
\bibitem[Jeffery et al.(1998)]{jef98} Jeffery, C. S., Hamill, P. J., Harrison, P. M.,
\& Jeffers, S. V. 1998, \aap, 340, 476
\bibitem[Jeffery, Hill \& Heber(1999)]{jeff99} Jeffery, C. S., Hill, P. W., \& Heber, U. 1999, \aap, 346, 491
\bibitem[Jeffery, Woolf \& Pollacco(2001)]{jeff2001} Jeffery, C. S., Woolf, V. M.,
\& Pollacco, D. L. 2001, \aap, 376, 497
\bibitem[Jeffery(2008)]{jef08} Jeffery C. S. 2008, ASP Conf. Ser. 391, 53
\bibitem[Karakas(2010)]{karakas2010a} Karakas, A. I. 2010, \mnras, 403, 1413
\bibitem[Karakas, Campbell \& Stancliffe(2010)]{karakas2010b} Karakas, A. I., Campbell, S. W.,
Stancliffe, R. J. 2010, \apj, 713, 374
%\bibitem[Kurucz(1998)]{kur98} Kurucz, R. L. 1998, http://cfaku5.harvard.edu
\bibitem[Lambert \& Rao(1994)]{lambrao94} Lambert, D. L., \& Rao, N. K. 1994,
JApA, 14, 47
\bibitem[Lanz \& Hubeny(2003)]{lh03} Lanz, T., \& Hubeny, I. 2003, \apjs, 146, 417
\bibitem[Lanz \& Hubeny(2007)]{lh07} Lanz, T., \& Hubeny, I. 2007, \apjs, 169, 83
%\bibitem[Lodders(2003)]{lod2003} Lodders, K. 2003, \apj, 591, 1220
\bibitem[McCarthy et al.(1993)]{McCar93} McCarthy, J. K., Sandiford, B. A.,
Boyd, D., \& Booth, J. 1993, \pasp, 310, 881
\bibitem[Morel \& Butler(2008)]{mandb08} Morel, T., \& Butler, K. 2008, \aap, 487, 307
\bibitem[Nieva \& Przybilla(2008)]{nieva2008} Nieva, M. F., \& Przybilla, N. 2008,
\aap, 481, 199
\bibitem[Pandey et al.(2001)]{pan2001} Pandey, G., Rao, N. K., Lambert, D. L.,
Jeffery, C. S., \& Asplund, M. 2001, \mnras, 324, 937
\bibitem[Pandey et al.(2006)]{pan2006} Pandey, G., Lambert, D. L.,
Jeffery, C. S., \& Rao, N. K. 2006, \apj, 638, 454
\bibitem[Pandey \& Reddy(2006)]{panred2006} Pandey, G., \& Reddy, B. E. 2006,
\mnras, 369, 1677
\bibitem[Pandey(2006)]{panF2006} Pandey, G. 2006, \apjl, 648, L143
\bibitem[Pandey, Lambert \& Rao(2008)]{pan08} Pandey, G., Lambert, D. L.,
\& Rao, N. K. 2008, \apj, 674, 1068
\bibitem[Popper(1942)]{pop1942} Popper, D. M. 1942, \pasp, 54, 160
\bibitem[Rao \& Lambert(1996)]{raolamb96} Rao, N. K., \& Lambert, D. L. 1996,
ASP Conf. Ser. 96, 43
\bibitem[Rao et al.(2004)]{rao04} Rao, N. K., Sriram, S., Gabriel, F., Prasad,
B. R., Samson, J. P. A., Jayakumar, K., Srinivasan, R., Mahesh, P. K.,
\& Giridhar, S. 2004, Asian Journal of Physics, 13, 267
\bibitem[Rao et al.(2005)]{rao05b} Rao, N. K., Sriram, S., Jayakumar, K.,
\& Gabriel, F. 2005, JAA, 26, 331
\bibitem[Rao \& Lambert(2008)]{raolamb2008} Rao, N. K., \& Lambert, D. L.  2008,
\mnras, 384, 477
\bibitem[Ryde \& Lambert(2004)]{rydelamb2004} Ryde, N., \& Lambert, D. L.,
2004, \aap, 415, 559
\bibitem[Saio \& Jeffery(2002)]{saio02} Saio, H., \& Jeffery, C. S. 2002, \mnras, 333, 121
\bibitem[Sch\"{o}nberner(1979)]{sch79} Sch\"{o}nberner, D. 1979, \aap, 79, 108
\bibitem[Seaton(1998)]{seat98} Seaton, M. J. 1998, J. Phys. B: At. Mol. Opt. Phys, 31, 5315.
%\bibitem[Sigut(1999)]{sig99} Sigut, T. A. A. 1999, \apj, 519, 303
\bibitem[Thackeray \& Wesselink(1952)]{thack1952} Thackeray, A. D.,
\& Wesselink, A. J. 1952, The Observatory, 72, 248
\bibitem[Tull et al.(1995)]{tull95} Tull, R. G., MacQueen P. J., Sneden, C., \&
Lambert, D. L. 1995, \pasp, 107, 251
\bibitem[Webbink(1984)]{webb84} Webbink, R. F. 1984, \apj, 277, 355
\bibitem[Werner \& Herwig(2006)]{werner2006} Werner, K., \& Herwig, F. 2006,
\pasp, 118, 183
\bibitem[Werner et al.(2008)]{werner2008} Werner, K., Rauch, T., Reiff, E.,
\& Kruk, J. W. 2008, ASP Conf. Ser. 391, 109
\bibitem[Wiese, Smith, \& Glennon(1966)]{wiese66} Wiese, W. L., Smith, M. W.,
\& Glennon, B. M. 1966, NSRDS--NBS 4

\end{thebibliography}
\end{document}